\documentclass{article}

\usepackage{inconsolata}
\newcommand{\asm}{\texttt}

\usepackage{multirow}
\usepackage{tabularx}
\usepackage{hyperref}
\usepackage{framed}
\usepackage{color}
\usepackage{xcolor}
\usepackage{xspace}
\usepackage{amsmath}
\usepackage{amssymb}
\usepackage{arydshln}
\usepackage{makecell}
\usepackage{amsmath}
\usepackage{xcolor}
\usepackage{textcomp}
\usepackage{booktabs}
\usepackage{float}
\usepackage{times}
\usepackage{mdframed}
\usepackage{latexsym}
\usepackage{graphicx}
\usepackage{wrapfig}
\usepackage{enumitem}
\usepackage{url}
\usepackage{stfloats}
\usepackage[caption=false]{subfig}

\usepackage{xcolor}
\usepackage{pifont}

\usepackage{authblk}
\usepackage[left=3cm, right=3cm]{geometry}

\usepackage{float}

\usepackage{url}

\definecolor{mGreen}{rgb}{0,0.6,0}
\definecolor{mGray}{rgb}{0.5,0.5,0.5}
\definecolor{mPurple}{rgb}{0.58,0,0.82}
\definecolor{backgroundColour}{rgb}{0.94, 0.97, 1.0}

\begin{document}
\title{BinBert: Binary Code Understanding with a Fine-tunable and Execution-aware Transformer}

\author[1]{Fiorella Artuso}
\author[1]{Marco Mormando}
\author[1]{Giuseppe Antonio Di Luna}
\author[1]{Leonardo Querzoni}
\affil[1]{Sapienza University of Rome}

\maketitle

\begin{abstract}
A recent trend in binary code analysis promotes the use of neural solutions based on instruction embedding models.  
An instruction embedding model is a neural network that transforms sequences of assembly instructions into embedding vectors. 
If the embedding network is trained such that the translation from code to vectors partially preserves the semantic, the network effectively represents an {\em assembly code model}.

In this paper we present BinBert, a novel assembly code model. BinBert is built on a transformer pre-trained on a huge dataset of both assembly instruction sequences and symbolic execution information. BinBert can be applied to assembly instructions sequences and it is {\em fine-tunable}, i.e. it can be re-trained as part of a neural architecture on task-specific data. Through fine-tuning, BinBert learns how to apply the general knowledge acquired with pre-training to the specific task. 

We evaluated BinBert on a multi-task benchmark that we specifically designed to test the understanding of assembly code. The benchmark is composed of several tasks, some taken from the literature, and a few novel tasks that we designed, with a mix of intrinsic and downstream tasks. 

Our results show that BinBert outperforms state-of-the-art models for binary instruction embedding, raising the bar for binary code understanding.

\end{abstract}

% !TEX root =  main.tex
\section{Introduction}
\label{sec:introduction}

A recent body of works in the literature has shown that Deep Neural Networks (DNNs) can successfully solve several binary analysis tasks. DNNs today show state of the art performances for binary similarity \cite{safe19mass,  nmt19Zuo, orders20Yu}, compiler provenance \cite{Himalaya18Chen, InvestigatingGE19mass, IdentifyingCA20Pizzolotto}, function boundaries detection \cite{RecognizingFB15Shin}, decompiling \cite{coda19fu}, automatic function naming \cite{nero20david, XFL21evans} and others.

DNN designers must decide how to feed binary code to their models. One possibility is to use manually-identified features. This approach requires a domain expert which identifies features of interest forecasting their helpfulness in solving the task at hand. This approach is known to produce problem-specific features and injects a human bias inside the system. Recent solutions automatically transform binary code into a representation usable by the neural network layers.

A common technique is to transform assembly instructions into representational {\em embeddings vectors}, similarly to what has been done in the Natural Language Processing (NLP) field with the word embedding revolution \cite{w2v}. Several works \cite{safe19mass,  nmt19Zuo, eklavya, palmtree2021li} proposed refined techniques to transform a single instruction into a vector of real numbers while capturing its semantic (e.g. all vectors of arithmetic instructions are clustered in the vector space). By using this approach, sequences of instructions are transformed into sequences of fixed-size vectors that can be fed into standard DNNs.

\subsection{Execution-aware Binary Code Interpretation}
A common weakness of all these solutions is {\em the lack of context}: an instruction is always represented by the same vector, irrespectively of where it appears. However, the semantic of a single assembly instruction is strongly limited (more than a word in natural language), and non-trivial concepts in assembly code are almost always encoded by a sequence of instructions (e.g., loops, swap of variables in memory, calling conventions, etc). Complex semantics, that span sequences of several assembly instructions, are hardly representable if embeddings of instructions are created in {\em isolation}; they have to be learned by the neural architecture using the embeddings. 

Code is a means of communication between humans and machines, thus, unlike natural language, it has a dual nature. One level represents the syntactic and semantic meaning that can be inferred from its {\em static form}; the other level is the possibility of being {\em executed}. Some aspects of an Instruction Set Architecture (ISA) can be easily appreciated only when code is executed (e.g., the dependencies introduced by \asm{RFLAGS} in X64). Moreover, semantically equivalent but syntactically different sequences of instructions are easily recognized when executed. 
Surprisingly, all embedding techniques we are aware of, only consider the static aspect of binary code.

\subsection{Fine-tunable Assembly Model}
\begin{figure*}[t!]
\center
    \includegraphics[scale=0.6]{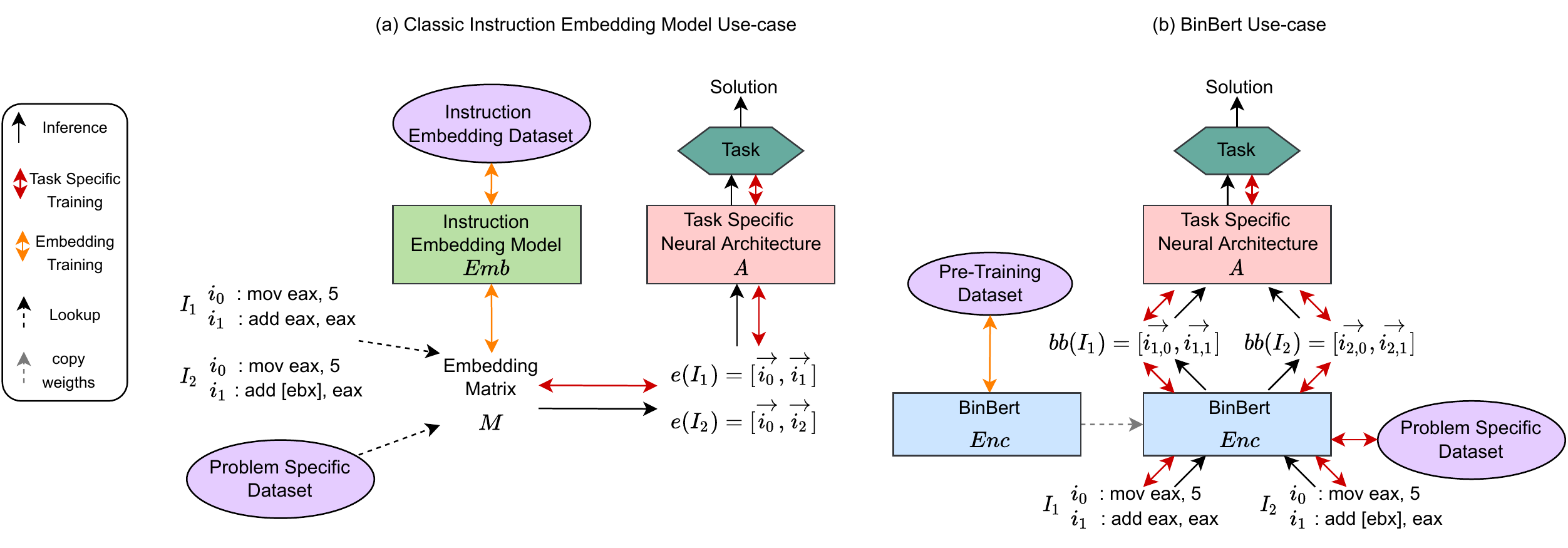}
    \caption{Comparison between an instruction embedding model (a) and the BinBert assembly model (b).  In (a) the embedding model trained on a corpus $C$ is a lookup mechanism that transforms instructions into vectors. Semantically relevant information have to be encoded in the embedding matrix $M$. The model cannot directly transfer knowledge into the final task. The BinBert assembly language model can use all the parameters of its neural architecture to store semantically relevant information that is then transferred to the downstream task with end-to-end training. \label{fig:instembvsbert}}
\end{figure*}

A second limitation of current solutions stems from the fact that all proposed models are non fine-tunable (see use-case (a) of Figure \ref{fig:instembvsbert}). The DNN creates an embedding vector for each instruction, and this vector is then fed into the specialized neural architecture $A$ that solves the target task. In this paradigm, there is no feedback from the specific problem to the embedding layer. All the semantically relevant information has to be stored into a single vector of real numbers.  It is clear that there is a limit on the information encodable in a small fixed-size vector.

We propose a different approach based on a {\em fine-tunable} assembly model (use-case (b) of Figure \ref{fig:instembvsbert}). This is a transformer-based encoder \cite{transformer} pre-trained on a large corpus of assembly code using specific tasks. During pre-training, the encoder learns a general semantic of assembly sequences that is context and execution aware. 
Then the pre-trained model is a part of a DNN $A$ that solves a specific {\em downstream task} (e.g., compiler provenance, function similarity, and others).  The DNN, including the assembly model, is retrained end-to-end on a small amount of problem-specific data during the fine-tuning process.
Intuitively, during pre-training, the encoder stores the learned information in its internal weights (tens of millions against the few hundreds of an embedding vector), while during fine-tuning, weights are rearranged and changed to apply acquired knowledge for the final goal (i.e., {\em knowledge transfer}). 
This paradigm is state of the art for NLP and works well also if the fine-tune dataset is small. This is useful whenever the definition of a labeled dataset requires expensive manual effort.

\subsection{Our proposal: BinBert}
In this paper we introduce {\em BinBert}, a fine-tunable assembly code model based on a transformer encoder that is execution-aware. 
To inject execution awareness into our model, our idea is to symbolically execute snippets of assembly code. Specifically, we use a symbolic execution engine that transforms sequences of assembly instructions connected by a data-dependency relationship (the strands introduced in \cite{statbinsim16david}) into sets of semantically equivalent symbolic expressions. These expressions are a functional representation of the input-output relationship of the strand. 
We designed a novel pre-training process that forces BinBert to learn the correct matching between an assembly sequence and an equivalent symbolic expression and to translate assembly code into symbolic expressions and vice-versa. 

We train BinBert on a new large dataset of assembly sequences and symbolic expressions derived from symbolic execution, obtaining a general-purpose assembly code model. We remark that symbolic execution is needed only in the pre-training phase; no code execution is required while using the model for inference tasks.

We tested BinBert on a multi-task benchmark for binary code understanding that we built. Tasks in the benchmark range from intrinsic ones, aimed at evaluating how the pre-trained BinBert captures the semantic of instructions and sequences, to extrinsic downstream tasks, in which we fine-tune BinBert for problems on assembly sequences and binary functions.  In all our experiments BinBert raises the performance bar outperforming the current state of the art (including the recent PalmTree \cite{palmtree2021li}) and specific solutions created for the binary similarity problem. 

In summary, this paper provides the following contributions:
\begin{itemize}
\item a novel training task that makes the training of an assembly code model execution-aware by using symbolic expressions derived from the symbolic executions of assembly snippets:
\item BinBert, a pre-trained execution-aware transformer model for X64, that can be plugged into DNNs for binary analysis;
\item the first multi-task benchmark designed to test the binary code understanding of assembly models. The benchmark is composed of well-known tasks selected from the literature for their relevance, and two novel tasks (strand recovery and execution) for the semantic understanding of assembly sequences; 
\item an in-depth performance evaluation of BinBert based on our benchmark that shows how execution awareness improves the performance of an assembly model. To the best of our knowledge, we are the first to thoroughly test the impact of the fine-tuning paradigm on assembly. We show that, as already shown in the NLP field, the pre-training/fine-tuning approach has a huge positive impact on all downstream tasks. As a consequence, BinBert outperforms the current state-of-the-art instruction embedding technique. 
\end{itemize}

%! Author = giuseppe
%! Date = 26/04/22
% !TEX root =  main.tex

\section{Background}\label{sec:back}
In this section, we introduce the general theoretical concepts behind the instruction embedding techniques and focus on the current state of the art. Afterwards, we detail weak points and gaps in current solutions, discussing how these influenced our proposal.

\subsection{Instruction Embedding Models}
\label{subsec:iem}

An instruction embedding model takes as input an assembly instruction $i$ from a vocabulary $V$ of size $d$ and it returns a vector of real numbers $e(i)=\vec{i} \in \mathbb{R}^{n}$, $n$ is the embedding size (typically $n \in [128,1024]$). The vector $\vec{i}$ is a {\em dense representation} of the instruction $i$.%, this terminology marks the difference with the old {\em one-hot encoding} approach in which the instruction is converted into a unit vector in $\mathbb{R}^{d}$. Dense representations reduce the curse of dimensionality associated with one-hot encoding.

In the simplest embedding scheme a random matrix $M$ of size $\mathbb{R}^{d \times n}$ is created, each instruction is mapped univocally to a row of $M$. With the matrix a sequence of instructions $I=[i_0,i_1,\ldots,i_{m}]$ is converted into a sequence of vectors $e(I)=[\vec{i_0},\vec{i_1},\ldots,\vec{i_{m}}]$ using a lookup mechanism.
This sequence is fed into the task-specific DNN $A$. The matrix $M$ is usually {\em trainable}: its elements are trainable weights and are modified during the training of $A$. %Sometimes also a fixed non-trainable matrix $M$ is used.

The groundbreaking idea of the embedding models is to generate the embedding matrix $M$ with a neural network $Emb$, formally speaking $M=Emb(A)$. The network $Emb$ is trained in an unsupervised way on a corpus $C$ of data. The use case of the instruction embedding models is reported in Figure \ref{fig:instembvsbert}.
 This corpora $C$ is composed of sequences of assembly instructions extracted from selected binaries.
Usually, the {\em distributed representation learning tasks} used by instruction embedding models are, apart from minimal modifications, the ones used in NLP by solutions such as word2vec \cite{w2v}, GloVe \cite{glove}, fastText \cite{fastext}, pv-dm \cite{pvdm}. The common goal is to train $Emb$ to produce an embedding vector that contains enough information to predict a masked instruction from its context in $C$. 
Most of the novelty of the instruction embedding system is in the preprocessing of instructions and in the definition of the assembly sequences composing $C$. 

\subsubsection{Preprocessing of Assembly Instructions} Assembly language and natural language are distinguished by the wide difference between the vocabulary size $d$. A natural language is usually composed of hundred thousands different words, while the number of possible distinct assembly instructions is much more. Consider the X64 ISA, a mov instruction can use $64$ bits to express immediates, offsets, and memory addresses, thus there can be $2^{64}$ different instructions that just move a value in a certain register. This makes raw assembly instructions impractical: a large vocabulary is discouraged \cite{chen-etal-2019-large} as it worsens the problem of out-of-vocabulary word (OOW) \cite{oov}.

Moreover, the exact value of an immediate is largely useless in a static analysis setting (e.g. a memory address of an unknown memory layout) \cite{safe19mass}. To ameliorate this problem, a lot of effort has been devoted to instructions preprocessing \cite{nmt19Zuo, deepbindiff, trex, asm2vec, palmtree2021li, safe19mass}. The standard of the field is to substitute all memory addresses and immediates above a certain threshold value with special symbols (e.g. \asm{IMM}).  Another design choice is whether to consider the entire assembly instructions as a token (used in \cite{safe19mass}), to split the assembly instructions into several tokens by separating opcodes and operands (used in \cite{asm2vec}) or to use a more fine-grained split strategy \cite{palmtree2021li}. Interestingly, no one used automatic tokenization such as WordPiece \cite{wordpiece} that are standard in NLP. 

\begin{figure}[t!]
    \center
    \includegraphics[scale=0.28]{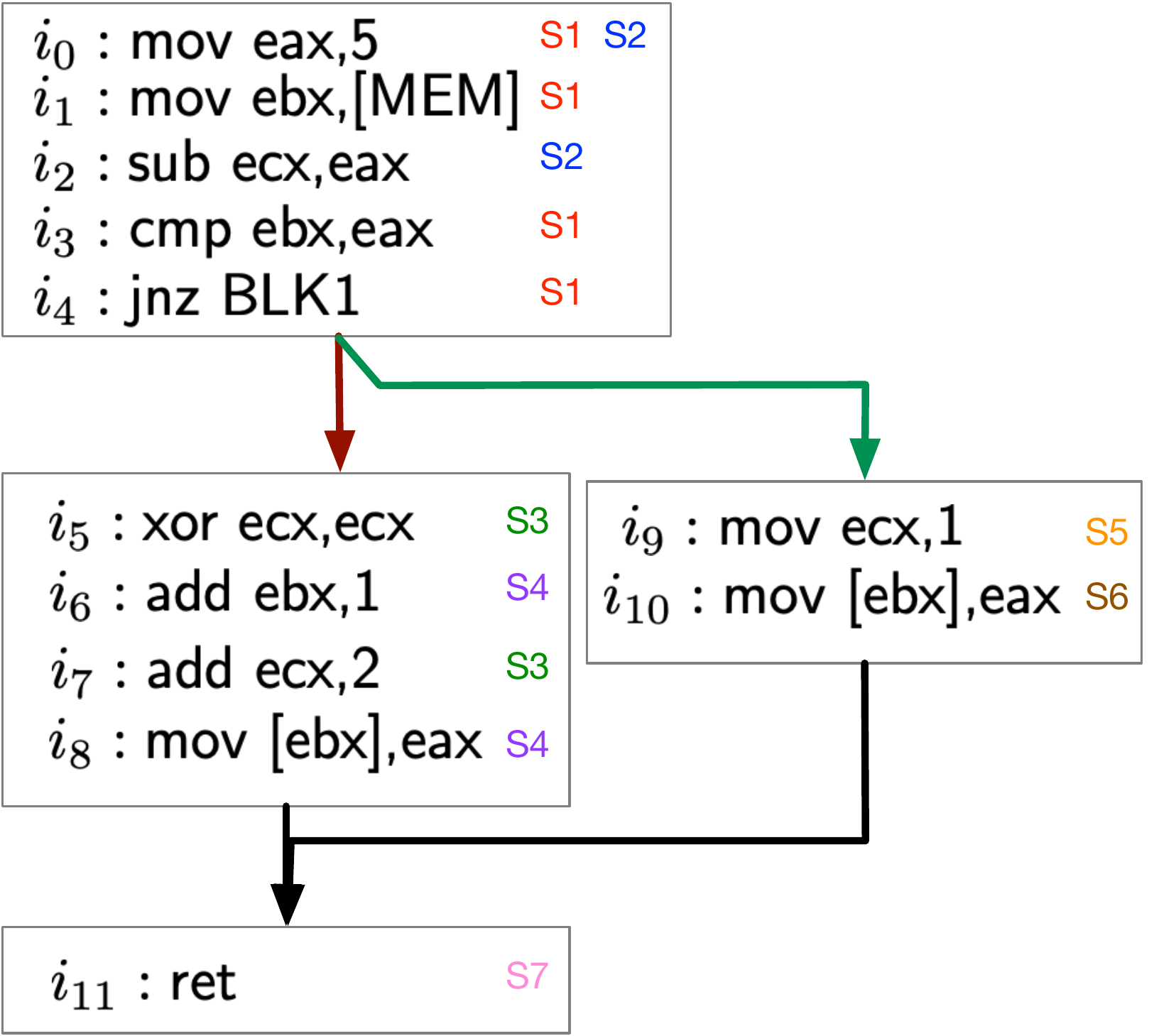}
    \caption{CFG of an imaginary function in x64 assembly. For each instruction we indicate with sx the block level strand to which it belongs. \label{cfgexample}}
\end{figure}

\subsubsection{Extraction of Assembly Sequences} \label{old:amsextract}
A key point is how to extract the sequences from the binary, as this defines the context in which an instruction appears. The context for instruction $i_x$ is composed by $k$ instructions appearing before/after $i_x$ in $C$. Used extraction strategies are:
\begin{itemize}
    \item Linearized CFG: In this case each sequence is a linearization of a CFG (commonly the one provided by a disassembler) \cite{safe19mass}. Blocks of the CFG that are not logically related could be sequentially placed in the linerization, and thus an instruction will see a noisy context. An example is the linearization of the CFG in Figure \ref{cfgexample} induced by instructions numbers: the context of instruction $i_8$ contains instructions $i_9,i_{10}$ that are not causally related. This inject noise in the learning process.
    \item CFG/ICGF: the sequences are extracted from the recovered CFG/ICFG. This is done either by using a random walk strategy (\cite{asm2vec}), or by taking as a sequence a single block \cite{nmt19Zuo}. The idea is to have a sequence that respects the logical control flow of the examined program, removing the source of noise highlighted in the previous strategy. We argue that this technique does not completely remove the presence of extraneous instructions in the context. Take the sequence of instructions $i_5,i_6,i_7,i_8$ in Figure \ref{cfgexample}, the context of instruction $i_7$ contains $i_6$ and $i_8$ that are not causally related.
\end{itemize}

\subsubsection{PalmTree}
\label{subsec:palmtree}
PalmTree  \cite{palmtree2021li} is an instruction embedding model that has shown state-of-the-art performances beating all the other embedding models on several tasks. PalmTree is based on a transformer model. PalmTree divides an instruction into tokens using a fine-grained strategy with manually made regexs. 
PalmTree is trained on pairs of instructions taken from the corpora $C$. PalmTree uses the standard MLM of Bert and two novel tasks: the CWP in which the network has to recognise if a pair of instructions is taken from the same context or not, and the DUP in which the network has to recognize if there is a data dependency between instructions. Once trained, PalmTree is used as an instruction embedding model (see use-case (a) in Figure \ref{fig:instembvsbert}): a sequence of instructions is embedded by applying separately the PalmTree model to each instruction. %(formally, representing PalmTree with the function $Emb$ given the sequence $I_0=[i_0,i_1,\ldots,i_m]$ the embeddings are $[\vec{i}_0=Emb(i_0),\vec{i}_1=Emb(i_1),\ldots,\vec{i}_m=Emb(i_m)]$).

\subsection{Weak Points and Gap Analysis}
We can now identify weak points of previous solutions and gaps in the current body of knowledge. From such analysis, we derive the research directions of our solution. 
 
\subsubsection{Focus on a Single Assembly Instruction and No Context-Awareness} Current solutions transform a single assembly instruction into a vector. This approach, despite its proven effectiveness \cite{eklavya, safe19mass}, has an implicit weakness: the semantic of a single assembly instruction is really limited. Even a perfect embedding model is constrained to produce embedding vectors that capture only the semantic of a single instruction. However, non-trivial concepts  are encoded using an entire sequence of instructions (e.g., swap of two variables in memory, spilling/restoring of variables in/from the stack,\ldots).

On the other hand, the meaning of an assembly instruction depends from its context. Consider the sequences $I_0:$ \asm{lea eax, [ebx * ecx + edx]; mov edi, eax} and $I_1:$ \asm{lea eax, [ebx * ecx + edx]; mov edi, [eax]}; in the first case the \asm{lea} is used to perform arithmetic operations (\asm{edi=} \asm{(ebx*ecx)+edx}), in the second case is used to compute a memory address.
An instruction embedding approach cannot encode these concepts, even if they are present in the corpora $C$. Therefore, (re)learning such patterns is demanded to the upward neural architecture $A$ (see Figure \ref{fig:instembvsbert}). This means that $A$ has to be trained on a large enough problem-specific dataset, even if $Emb$ was trained on large corpora of data.
Thus our first research direction is:
	
\begin{mdframed}[backgroundcolor=mGreen!8] 
 \small
{\bf RD 1}: Design an embedding model that could transform entire sequences of instructions into semantically representative embedding vectors. If the model is used to generate the embedding of an instruction, it has to take into account its context.
\end{mdframed}

\subsubsection{No Execution} The execution of assembly makes clear concepts that would be covert (or unavailable) from its static representation. Apart from the \asm{RFLAGS} example mentioned in the introduction, consider again the sequence $I_0:$ \asm{lea eax, [ebx * ecx + edx]; mov edi, eax} such sequence is semantically equivalent to  $I'_0:$ \asm{imul ebx, ecx; add ebx, edx; mov edi,eax}. The semantic equivalence could be easily discovered if information taken from the execution is inserted in the pre-training tasks. Unfortunately, current techniques totally neglect this aspect.  

\begin{mdframed}[backgroundcolor=mGreen!8] 
 \small
{\bf RD 2:} Design an embedding model that takes into account the execution of code. The impact of execution-related information has to be quantified with a specific ablation study.
\end{mdframed}

\subsubsection{No End-To-End Retraining (Fine-tuning)}
The use case of all the known instruction embedding models is the (a) of Figure \ref{fig:instembvsbert}. In the NLP field, such approach has been largely substituted by encoder models pre-trained on corpora $C$ and then fine-tuned on a specific task (use-case (b) of Figure \ref{fig:instembvsbert}). 
During fine-tuning, the pre-trained encoder $Enc$ is trained with the network $A$ end-to-end on a problem specific 
labeled dataset. The back-propagation algorithm optimizes the internal weights of network $A$ and $Enc$; this optimization modifies the weights of $Enc$ so that the knowledge learned during pre-training is applied to the downstream task. As a matter of fact, when $Enc$ is a transformer, $A$ is usually a linear classifier or another simple neural network, since most of the work is performed by $Enc$.
%There are several hypothesis on why in NLP the use of pretraining before a specific fine-tuning is beneficial \cite{}, as example it has been shown that the several layers of Bert during pretraining automatically learn the classic pipeline used in non-neural based NLP works \cite{}. 

%One would wonder what could happen when the same idea is applied to assembly language models. 
% We are not aware of any work that has focussed on the pretrain/fine tune paradigm for assembly sequences. 
No one has extensively studied how this paradigm copes with solving different goals on assembly language, goals selected to test the semantic and syntactic comprehension of assembly code. This interesting gap of the current body of knowledge gives us a new research direction:
\begin{mdframed}[backgroundcolor=mGreen!8] 
 \small
{\bf RD 3:} Design an embedding model that can be trained end-to-end on a specific task, transferring the general knowledge learned during pre-training. The model has to be evaluated on a multi-task benchmark designed to thoroughly test the syntactic and semantic understanding of the assembly language.
\end{mdframed}

%\subsubsection{Manual Tokenization of Assembly Instructions}
%
%	   \begin{mdframed}[backgroundcolor=mGreen!8] 
% \small
%{\bf RD 4:} 
%\end{mdframed}

% !TEX root =  main.tex
\section{The BinBert Solution}
In this section we describe BinBert. We first give an overview of the system, briefly describing the transformer architecture \cite{transformer}. We then give the details of the innovative aspects of BinBert.

\subsection{Overview}

The neural architecture of BinBert is the standard transformer encoder \cite{transformer} used by Bert \cite{bert}. A transformer encoder processes sequential data using an attention mechanism, which allows for both the creation of more informative embeddings (by focusing only on relevant parts of the sequence) and good performances (the attention mechanism is implemented using matrix multiplication that is highly parallelizable on GPUs). In detail, a transformer encoder is composed of $N$ identical layers stacked one on top of the other, where each layer consists of two sublayers: a multi-head self-attention mechanism and a fully connected feed-forward network. Practically speaking, a sequence of $n$ tokens is transformed into a sequence of $n$ latent vectors (one for each token) with a mechanism that we will explain in Section \ref{sec:bertnovel}; this sequence is fed into the initial layer of the encoder. Each other layer takes as input the hidden state token vectors returned by the previous layer. The output of the encoder is a sequence of $n+1$ embedding vectors: one for each input token and a special embedding for the entire sequence (the \asm{[CLS]} vector described in Section \ref{sec:bertnovel}).

To avoid the vocabulary inflation generated by the use of raw assembly instructions, BinBert substitutes memory addresses and immediates above a certain threshold with special symbols. Moreover, BinBert splits a single assembly instruction into several tokens using WordPiece \cite{wordpiece}.

In BinBert we decided to completely remove the noise given by instructions that are contextually related but have no logical relation (see Section \ref{old:amsextract}) by extracting sequences representing \emph{strands} \cite{statbinsim16david}. Strands are sequences of causally related instructions computing the values of a certain variable. In this way, the context of an instruction never contains extraneous instructions introduced by compiler optimizations.
We symbolically execute each strand to extract a set of symbolic expressions; these expressions will be used in our training tasks as a means to inject execution-related information into the pre-training.

During pre-training, BinBert learns the matching between symbolic expressions and strands (this is done using positive/negative pairs); at the same time, samples are partially masked according to a translation task.
%From a certain perspective, this is similar to training BinBert on two different languages, the assembly code and the symbolic expression. It is worth notice, that multi-lingual training of Bert is an hot topic in NLP, and it has been shown that training the model on two (or more) different languages increases its performance on single language tasks.

\subsection{Instructions Preprocessing and Assembly Sequences Extraction}
\label{sec:inst_pp}

We preprocess each assembly instruction substituting immediates above a threshold (5000 in our experiments) with the value \asm{IMM} (the same is done for offsets and memory addresses). We use the special symbol \asm{MEM} in case of jumps. We use a threshold-based approach as small immediate values are like to carry informative content (comparison with small constants in if and loops, PC/stack relative displacements that identify variables in memory).  All immediates/offsets are converted to decimal format.
For call instructions, we distinguish if the called function is user-defined or belongs to libc. For user-defined functions, we substitute the called address with \asm{func} (our system is usable on stripped binaries). If it is a call to libc, we substitute the address with the function name (e.g., \asm{call printf}), since external symbols cannot be stripped. Indirect calls are left untouched.

After preprocessing, each instruction is tokenized using WordPiece \cite{wordpiece}. The latter uses a probabilistic approach to learn how to tokenize instructions in a way that minimizes the vocabulary size and the OOW problem. Contrarily to manually made regexes, WordPiece automatically learns how to split complex opcodes (as an example \asm{cmovz} will be split in \asm{cmov} and \asm{z} helping the model in understanding the relationship between the cmovX family of X64). We use WordPiece also on symbolic expressions, This provides a uniform tokenization mechanism and vocabulary for the two distinct languages (asm/sym. expr.) used for BinBert.

\paragraph{Assembly Sequences - Strands extraction} \label{par:strandextraction}

In BinBert we use the concept of strands to extract the sequences of assembly instructions on which our model is trained. This does not mean that BinBert cannot be fine-tuned and used on CFG blocks or entire functions as our experiments will show.

A strand, originally defined in \cite{statbinsim16david}, is a slice of a CFG  block constituted by all the instructions that are connected by def-use dependences. More specifically, we consider as an output variable of a block a memory location or a register on which the last operation is a write or the check of a jump. Starting from this variable we construct a  {\em backward-slice} of the block including all the instructions from which the value of such variables depends. To make the concept clear, consider the example in Figure \ref{cfgexample}; the first block contains two strands S1 and S2: S1 is composed by instructions $i_0,i_1,i_3,i_4$ that influence the \asm{RFLAGS} register later checked by $i_4$; strand S2 is composed by $i_0,i_2$ which define the value of variable \asm{ecx}. Other examples of strands are in the figure. We enrich the original definition of strand, by considering as a single strand all the instructions that prepare the input values for a call. In this case, the strand will be constituted by all the aforementioned instructions and the call instruction itself.
Therefore, we build our training corpora $C$ by extracting the CFGs in binary, and then decomposing all the blocks in strands. The strands will be the basic sequences in $C$.

This decomposition has several advantages: it completely removes the noise introduced by instructions that co-occur in the same context only for compiler optimization reasons and the model leanrs the entire {\em ``causal context''} of an instruction so it is able to see long dependencies among instructions.
%Notice that using strands is an extension of the idea of sampling pairs from the DFG used by PalmTree. There is a key difference, in our case for each sample the model has the entire {\em ``causal context"} of an instruction, and it is able to see long dependencies that are otherwise covert (in PalmTree a single edge of DFG is considered).

\subsection{Symbolic Execution} \label{sec:symbex}
In BinBert, we use symbolic execution to convert each strand into a representative symbolic expression. The symbolic execution engine works on strands of assembly instructions before we apply the preprocessing (it needs the actual value of immediates and memory locations). The engine is built on angr \cite{angr}.  During the execution we consider as inputs the variables on which the first operation executed by the strand is a read, and as outputs the variables on which the last operation executed is a write.

Each time the strand writes on a variable (either a memory location or a register) we express the written value using a symbolic expression. When a variable is read we may have that either the variable is an input (no one wrote on it) or it contains a symbolic value. In case the variable is an input, we set its symbolic value to its address or register name (as example, for \asm{mov eax, [rbp+4]} we have \asm{eax=*(rbp+4)}, for \asm{mov eax, ebx} we have \asm{eax=ebx}).

The symbolic expression obtained with this process may have one of three possible forms:
 \begin{itemize}
 \item If the strand computes the value of a certain variable, the symbolic expression describes the value in the output variable as a function of the strand inputs. An example is in the first row of Table \ref{table:sym_expr}: we have the symbolic expression \asm{rcx=-1\,\, add\,\, (0\,\, Concat \,\,rsi[1:0])}. This expression is for the output variable \asm{ecx}: the \asm{and} operation extracts the two least significant bits from \asm{rsi}, the result is extended to 64 bits, and then decremented by the \asm{rep}. The expression only contains the extended register of X64, this is a design choice that we discuss later in Section \ref{sec:bertnovel}.

 \item If the strand computes the predicate checked by a conditional branch, then our symbolic expression will be the comparison of the jump condition with a symbolic expression of the value used in the predicate. As an example, consider the second row of Table \ref{table:sym_expr}; in this case, the symbolic expression is compared with $0$ using the not equal (ne) predicate.

\item Finally, in case our strand computes the arguments used by a call instruction, our symbolic expression will be the call to the specific function (including the symbolic name if it is a libc call), where all the arguments are substituted by the symbolic expressions of their values. We extracted function arguments following the X64 calling convention. An example of a call expression is in the third row of Table \ref{table:sym_expr}. 
\end{itemize}
From a single strand, we may have several symbolic expressions. For instance, in the first row of Table \ref{table:sym_expr} the \asm{rep stosb} instruction, repeatedly puts the content of \asm{al} into the memory pointed by \asm{rdi} for \asm{ecx} times by decrementing \asm{ecx} and incrementing \asm{rdi} at each iteration. This means that the strand has three output variables (a memory location and two registers), and each one will have its symbolic expression. We will call such set of symbolic expressions the {\em representative set} of the strand. 

% \usepackage{multirow}
% \usepackage{graphicx}
\begin{table}[]
\resizebox{\linewidth}{!}{%
\begin{tabular}{|l|l|}
\hline
\textbf{Strands} & \textbf{Symbolic Expressions} \\ \hline
\multirow{3}{*}{
	\begin{tabular}[c]{@{}l@{}}
		\asm{mov ecx, esi}\\ \asm{and ecx, 3}\\ \asm{rep stosb byte ptr {[}rdi{]}, al}
	\end{tabular}
}
& \asm{rcx = -1 add ( 0 Concat rsi{[}1:0{]} )} \\ \cline{2-2}
& \asm{rdi = 1 add rdi}                                      \\ \cline{2-2}
& \asm{*(rdi) = al}                                             \\ \hline

\begin{tabular}[c]{@{}l@{}}
	\asm{mov rax, qword ptr {[}rbp - 168{]}} \\ \asm{mov eax, dword ptr {[}rax + 24{]}} \\ \asm{test al, 2}\\ \asm{jne MEM}
\end{tabular}
& \asm{0 Concat *(*(rbp add -168) add 24){[}1:1{]}  ne 0} \\ \hline

\begin{tabular}[c]{@{}l@{}}
	\asm{mov esi, IMM}\\ \asm{mov rdi, qword ptr {[}rbp{]}}\\ \asm{call fprintf}
\end{tabular}
& \asm{fprintf(*(rbp), IMM)}                                               \\ \hline

\end{tabular}%
}
\caption{Examples of symbolic expressions obtained by different strands.}
\label{table:sym_expr}
\end{table}

 \paragraph{Preprocessing and Tokenization of Symbolic Expressions}
 The symbolic expressions of each strand are preprocessed similarly to assembly. Large numerical constants are substituted with the special symbol \asm{IMM}, for floating point numbers all the digits after two decimals are truncated.  The symbolic expressions are tokenized using WordPiece. During the preprocessing phase, we also substitute the name of all registers to their extended form (i.e., we use \asm{rax} instead of \asm{eax}), we do so to help the network in understanding the relationships between names used to address different parts of the same logical register; this step is not applied while preprocessing assembly instructions (i.e., we leave \asm{eax} in the strand).

\subsection{BinBert Input Representation and Pre-Training Tasks} \label{sec:bertnovel}
BinBert is fed with pairs $<$strand,symbolic expression$>$ and pre-trained on two tasks that we name Execution Language Modeling (ELM) and Strand-Symbolic Mapping (SSM).
\begin{figure}[t!]
\center
    \includegraphics[width=\linewidth]{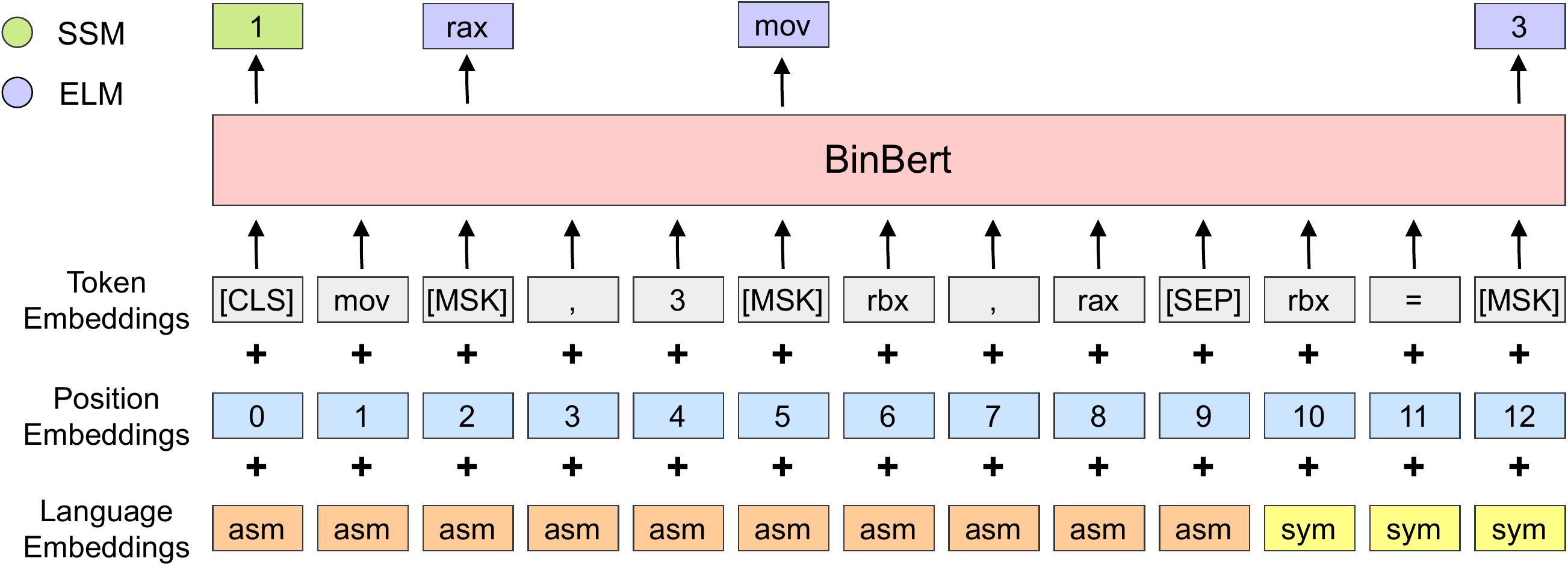}
    \caption{Example of BinBert pre-trained on the Execution Language Modeling (ELM) and on the Strand-Symbolic Mapping (SSM) tasks.
\label{fig:binbert}}
\end{figure}

\paragraph{Input Representation}\label{sec:inputrep}
A BinBert input is a tokenized strand-symbolic expression pair. Two special tokens are added to each sample: \asm{[SEP]} used to keep the distinction between assembly and symbolic expression,  \asm{[CLS]} prepended to all the samples. The hidden state of the  \asm{[CLS]} token in the last hidden layer is usually used to obtain a latent vector representation of the whole sequence \cite{bert} (see Figure \ref{fig:binbert}).

We use dynamic padding, thus shorter sequences are padded with the special token \asm{[PAD]} and sequences longer than a threshold (we use 512 in our experiments) are truncated. Prior to the transformer architecture, each token is converted into a vector by using the lookup mechanism described in Section \ref{subsec:iem}. Then, token embeddings are summed up with both position and language embeddings \cite{bert, XLM}. Position embeddings are used to make the model aware of the notion of sequences, while language embeddings are used to distinguish assembly from symbolic expressions (see Figure \ref{fig:binbert}).

\paragraph{Pre-Training Tasks}\label{sec:pretrain}
The first task is \textbf{Execution Language Modeling} (ELM). The goal of ELM is to mask a certain percentage $mp$ of tokens in the input pairs and to make the network predict the original ones. As in Bert, the tokens that have to be predicted are either substituted with a special token \asm{[MASK]}, replaced with a random word, or left untouched with 80-10-10 probabilities respectively.  Figure \ref{fig:binbert}, shows an example in which the tokens \asm{rax}, \asm{mov}, and \asm{3} are masked and the network tries to reconstruct them as output. Notice that, in order to reconstruct a token contained in a strand, the network needs to pay attention to both the strand and its corresponding symbolic expression (the same holds for a token in the symbolic expression). This means that in order to predict the value \asm{3} for register \asm{rbx} the network is forced to understand the data flow in the strand, thus making the semantic of each strand instruction explicit. As in \cite{bert} a linear layer is added on top of the last hidden layer, in particular to the hidden states of the masked tokens; this layer will guess the masked tokens.

%This task is similar to the Translation Language Modeling (TLM) task of \cite{XLM}, in which two parallel sentences in two different languages are fed to Bert and the masking is applied to both sentences. 

Mathematically speaking, we have a dataset $\mathcal{D}$ of $<$strand,symbolic expression$>$ pairs $I=[t_1,\ldots,t_n] \in \mathcal{D}$ where each token $t_i$ belongs to a vocabulary $V \in \mathbb{R}^{d}$. We feed each pair to BinBert to obtain context-aware hidden vectors $H=Binbert(I)=[\vec{h_1},\ldots,\vec{h_n}]$ as output. Now, consider a function $msk(I,m_p)$ which randomly mask $m_p$ percentage of tokens of $I$, then the goal of the network is to predict the probability that a token $t_i$ corresponds to the target word $\hat{t}_i$ with the softmax function:
\begin{equation}
p(t_i=\hat{t}_i| I) = \frac{e^{\vec{w}_{\hat{t}_i} \cdot \vec{h_i}}}{\sum_{k=1}^{d} e^{\vec{w}_{\hat{t}_k} \cdot \vec{h_i}}}
\end{equation}
where $\vec{w}_{\hat{t}_i}$ is the weight vector of the linear layer for word $\hat{t}_i$. The loss of the ELM task is the cross entropy loss:

\begin{equation}
\mathcal{L}_{ELM}= - \sum_{I \in \mathcal{D}} \sum_{t_i \in msk(I,m_p)} \log p(t_i=\hat{t}_i| I) 
\end{equation}

The second task is the \textbf{Strand-Symbolic Mapping} (SSM), in which, the goal is to predict whether the symbolic expression is from the set of expressions representative of the strand (see Section \ref{sec:symbex}). To solve this task, we create both negative pairs by associating a strand with a random symbolic expression and positive pairs in which the symbolic expression is taken from its representative set. The ratio between positive and negative pairs is 50:50. An example of positive pair can be seen in Figure \ref{fig:binbert}: the input strand computes the value $3$ for register $rbx$ as stated by the corresponding symbolic expression. We believe that, with this task, the network is forced to learn the matching assembly snippets and symbolic expressions. This task is built by using a linear layer on top of the hidden state of the \asm{[CLS]} token in the last layer that will be used to classify a pair as negative or positive. In mathematical terms, the goal of this task is to evaluate the probability that the output label is one, i.e. the symbolic expression correctly computes a value in the strand:
\begin{equation}
p( y=1 | I) = \frac{e^{\vec{w_1} \cdot \vec{h}_{[CLS]}}}{ e^{\vec{w_0} \cdot \vec{h}_{[CLS]}} + e^{\vec{w_1} \cdot \vec{h}_{[CLS]}}}
\end{equation}
where $\vec{w_0}$ and $\vec{w_1}$ are the weight vector of the linear layer for label $0$ (negative pair) and $1$ (positive pair) respectively. The loss $\mathcal{L}_{SSM}$ of the SSM task is the standard cross entropy loss.

%\begin{equation}
%\mathcal{L}_{SSM}= - \sum_{I \in \mathcal{D}} \log p( y=1 | I)
%\end{equation}

The final loss on which BinBert is trained is the sum of the losses of the two tasks described above:

\begin{equation}
\mathcal{L}= \mathcal{L}_{ELM} + \mathcal{L}_{SSM}
\end{equation}

%We argue that feeding the network with strand-symbolic execution pairs rather than just assembly sequences enhances the performance of the overall model. A similar result is achieved in the NLP field, in which a cross-lingual model trained on parallel sentences in different languages,  enhances the performance of downstream applications with respect to monolingual models  \cite{XLM}.
%! Author = giuseppe
%! Date = 26/04/22
% !TEX root =  main.tex

\section{Evaluation Tasks}
When proposing a new assembly model an extensive experimental evaluation is fundamental. 
In NLP is customary to use standard multi-task benchmarks to evaluate language models \cite{glue}. Similar benchmarks for binary code do not exist. Therefore, we designed our benchmark by selecting several tasks. For each task defined on a sequence of assembly instructions, we have a version on strands and one on CFGs' blocks. This tests BinBert on sequences that are not strands. 

\subsection{Intrinsic Tasks}
The instrinsic tasks \cite{camacho-navigli-2016-find} directly use the embeddings produced by BinBert, there is no fine-tuning and the embeddings are not used as input for other models. 
\subsubsection{Opcode and Operand Outliers}
\label{subsec:outlier_detection}
In the {\bf opcode outlier} task we are given a set of five instructions. Four of these instructions belong to the same semantic class, and one is an outlier. As example, if the set is \{\asm{add eax, ebx; sub ebx, ecx; imul ecx, edx; add eax,5; call printf;}\}, the last one is an outlier. The network has to predict which is the outlier, among the 5 instructions.%To solve the task, we embed each instruction in the set and we evaluate if the embeddings are able to distinguish the outlier; this is done by computing the distance of each embedded instruction from the others and by predicting as outlier the most distant.
%We built opcode categories in the same way as PalmTree, by following the instruction categorization provided by the x86 Assembly Language Reference Manual \footnote{\url{https://docs.oracle.com/cd/E26502_01/html/E28388/ennbz.html}}.

The {\bf operand outlier} task is analogous, but in this case, the operands define the outlier. Given the set \{\asm{sub [eax+5], ebx; sub ebx, ecx; imul ecx, edx; add eax,ebx, add ebp,esp;}\}, the outlier is the first instruction: the only using a memory operand. Again, we followed the operand categorization provided by \cite{palmtree2021li}. %We selected them since they are tasks at instruction level.

\subsubsection{Strand and Block Similarity}
In the {\bf strand similarity} task we compute the embeddings of given strands with BinBert and use them to discover semantically similar strands. Two strands are similar if they have an overlapping semantic (non-empty intersection of the representative sets). More formally, we have a lookup database of $n$ strands $\mathbb{A}=\{a_1, \ldots, a_n\}$ and a query strand $q$. The lookup contains strands that are similar to $q$ and strands that are dissimilar. Given a number $k$ the network has to return the $k$ strands in $\mathbb{A}$ that are most similar to $q$. 
The {\bf block similarity} is analogous but done at basic block level. In this case, we use the definition of similar blocks used in \cite{deepbindiff}, where two blocks are similar if the DWARF information says so.

\subsection{Extrinsic Tasks}
In the extrinsic tasks, BinBert will be used as the encoding layer of a neural architecture and fine-tuned end-to-end. 

\subsubsection{ Strand and Block Similarity}
The {\bf extrinsic strand similarity} and {\bf extrinsic block similarity} are the same tasks of their intrinsic versions; in this case, we fine-tune BinBert using a dataset of similar and dissimilar pairs of samples. We decide to include these tasks as they will quantify the effect of fine-tuning on the creation of semantic preserving embeddings. These tasks are the dual of the compiler provenance below. 

\subsubsection{Strand and Block Compiler Provenance} In the {\bf strand compiler provenance} task the architecture has to recognise the compiler and the optimization levels used to generate a particular strand. This task has been previously proposed on functions \cite{InvestigatingGE19mass} and fragments of code \cite{oglassesx}. In compiler provenance, the network has to recognize the syntactic signature that a compiler, or optimization level, produces while in the strand and block similarity it has to abstract from such differences to get the meaning of the sequences. The {\bf block compiler provenance} task is analogous but at block level. 

\subsubsection{Strand Recovery and Execution}
We designed two novel tasks that test the semantic understanding of assembly sequences. In {\bf strand recovery} we provide to the network a basic block of the CFG where one instruction is marked. The DNN has to recognize all instructions in the same strand of the marked instruction. This task tests the understanding of the inputs/outputs of instructions, the network has to infer the dependency created by implicit registers such as RFLAGS.

In {\bf strand execution} a strand and a question are given to the network. The question is composed of an assignment for the inputs and a marked output. The network has to predict the value of the marked output. This task is interesting as it forces the network to concretely execute the snippet of assembly code and compute the correct output. 

\subsubsection{Function Level - Compiler Provenance and Similarity} 
Finally, our multitask benchmark contains two tasks at the function level. In the {\bf function compiler provenance} task the network is given an entire binary function and it has to predict the compiler used to generate the function and the optimization level. The {\bf function similarity} tasks is analogous to the extrinsic block similarity: given a database of functions and a set of function queries, for each query, the network has to return the similar functions in the dataset. We use the standard definition of function similarity \cite{safe19mass}: two assembly functions are similar when they derive from the same source code compiled with different compilers or optimization levels.  Function similarity is a current hot topic in research \cite{safe19mass,nmt19Zuo,deepbindiff,orders20Yu,trex} for its security implications.

%! Author = giuseppe
%! Date = 02/05/22
% !TEX root =  main.tex
\section{Datasets, Pre-Training and Implementation Details}
 
\subsubsection{Datasets}
We used two datasets, a pre-train dataset PTData used to pre-train BinBert, and a test dataset TestData for the downstream tasks. Our datasets are for X64. 
\paragraph{PTData - Pre-train Dataset} The dataset contains the projects:  ccv-0.7, binutils-2.30, valgrind-3.13.0, libhttpd-2.0, openssl-1.1.1-pre8, openmpi-3.1.1, coreutils-8.29, gsl-2.5, gdb-8.2, postgresql-10.4, ffmpeg-4.0.2, curl-7.61.0. The projects are compiled using: clang-3.8, clang-3.9, clang-4.0, clang-5.0, gcc-3.4, gcc-4.7, gcc-4.8, gcc-4.9, gcc-5.0.  For each compiler we compiled each project four times, one for each optimization level in $\{O0,O1,O2,O3\}$. 
We use radare2 (version 5.6.0) to extract functions signatures. We use angr (version 9.1.11611) to get CFGs, basic blocks, and strands and to obtain from each strand the set of symbolic expressions. 
After the removal of duplicates, we obtained $17.215.046$ pairs in the form $(strand, simexpr)$ as dataset for the SSM task. 

\paragraph{TestData - Test Dataset} The test dataset is obtained from diffutils-3.7, findutils-4.7.0, inetutils-2.0, mailutils-3.10, wget-1.20.3. We use clang-3.8, clang-6.0, clang-9, gcc-5, gcc-7, gcc-9, icc-21 and the 4 optimization levels  $\{O0,O1,O2,O3\}$ to obtain the raw binaries.
We will use these raw binaries to create the specific dataset for each task. Since some operations are task-dependent we discuss the specific split and format of the data in the experimental section of each task. We took care of removing duplicates so that the same sample will not be in fine-tune and test split.

\subsubsection{Model Parameters, Pre-Training and Implementation Details}

Our model is built using python 3, with pytorch (version 1.10.2+cu113) \cite{pytorch} and hugginface (version 4.16.02) \cite{huggingface}. We trained it on a DGX A100, using 4 A100 GPUs. 

\paragraph{Model and Pre-Training Parameters}
BinBert parameters are: sequence length $512$, hidden size $768$, intermediate size $3072$, $12$ attention heads and layers. We used AdamOptimizer and learning rate $0.0001$. The masking rate $mp$ is 0.3. The batch size for each device is $32$ with two steps of gradient accumulation; having $4$ GPUs the equivalent batch size is $256$.  We trained for $1$ epoch using $1425$ steps of warmup.
%Since each downstream task uses a different fine-tune setting, we discuss the specific of each in the relative experimental section.

% !TEX root =  main.tex

\section{Experimental Evaluation}
In our evaluation we answer the following experimental questions:

\begin{mdframed}[backgroundcolor=mPurple!8] 
 
 \begin{description}[style=unboxed,leftmargin=0cm]
\item[\bf RQ 1] Is an execution-aware transformer model trained on strands of assembly instructions the state-of-the-art assembly model for binary understanding?
\item[\bf RQ 2] What is the impact of pre-train on the performance across several binary understanding downstream tasks?
\item[\bf RQ 3] What is the impact of using an execution-aware pre-training?
\end{description}

\end{mdframed}

To answer {\bf RQ 1} we compare BinBert with PalmTree on all the tasks of our benchmark. We also compare BinBert with state-of-the-art function similarity solutions Safe \cite{safe19mass} and asm2vec \cite{asm2vec} on the extrinsic function similarity task. For a fair comparison, we retrain both PalmTree and asm2vec on our pre-train dataset, by using the same parameters of the original papers. %We do not retrain Safe \cite{safe19mass} since it was trained on the same dataset except for some compiler versions. 
On all the extrinsic tasks we fine-tune PalmTree \footnote{PalmTree authors highlighted the possibility of fine-tuning their system but they did not explore this possibility in their paper.} on exactly the same dataset we used to fine-tune BinBert. 
We are interested in assessing the effective contribution of the symbolic expressions used during pre-training ({\bf RQ 3}) and the pre-training itself on downstream applications ({\bf RQ 2}). To do so we will use the following baselines:
\begin{itemize}
\item \textbf{BinBert-MLM}: it is a model pre-trained on strands only with the standard Masked Language Modeling Task (MLM). In MLM only the assembly of a strand is given to the transformer during pre-training, the pre-training task is to recover masked tokens. The gap between this model and BinBert quantifies the impact of execution-awareness. 
\item \textbf{BinBert-FS}: it is a transformer encoder with the application-specific neural architecture trained from scratch on the specific downstream task. The gap between this model and BinBert quantifies the impact of pre-training. 
\item \textbf{PalmTree-FS}: similar to BinBert-FS but for PalmTree. The gap between this model and PalmTree quantifies the impact of pre-training. 
\end{itemize}

%feeding the transformer encoder with strands only and by pre-training the network simply on the Masked Language Modeling Task (MLM). We name the resulted baseline model BinBert-MLM. In the case of extrinsic tasks, we were also interested in evaluating the impact of pre-training on the final performances. To do so, instead of loading the pre-trained BinBert weights, we trained the transformer encoder with the application-specific heads from scratch. We will refer to such a model as BinBert-FS.
For the fine-tuned models we use the notation: \textbf{BinBert-FT}, \textbf{BinBert-MLM-FT}, and \textbf{PalmTree-FT}.

For each task, we will provide the details of the dataset, the solution we employed to solve the problem, the metrics used to evaluate the performance, and the final results.

\subsection{Intrinsic Tasks}\label{sec:it}
The results for intrinsic tasks are reported in Table \ref{table:intrinsic_res}.

\begin{table*}
%\resizebox{\textwidth}{!}{%
\center
\begin{tabular}{|ccc|ccccccccc|}
\hline
\multicolumn{3}{|c|}{\multirow{3}{*}{}}                                                                                                                                                                                       & \multicolumn{9}{c|}{\textbf{Models}}                                                                                                                                                                                                                                                                                     \\ \cline{4-12} 
\multicolumn{3}{|c|}{}                                                                                                                                                                                                        & \multicolumn{3}{c|}{\textbf{BinBert}}                                                                         & \multicolumn{3}{c|}{\textbf{BinBert-MLM}}                                                                     & \multicolumn{3}{c|}{\textbf{PalmTree}}                                                   \\ \cline{4-12} 
\multicolumn{3}{|c|}{}                                                                                                                                                                                                        & \multicolumn{3}{c|}{\textit{Accuracy}}                                                                        & \multicolumn{3}{c|}{\textit{Accuracy}}                                                                        & \multicolumn{3}{c|}{\textit{Accuracy}}                                                   \\ \hline
\multicolumn{1}{|c|}{\multirow{2}{*}{\textbf{\begin{tabular}[c]{@{}c@{}}Outlier\\ Detection\end{tabular}}}} & \multicolumn{2}{c|}{\textbf{Opcode}}                                                                            & \multicolumn{3}{c|}{\textbf{83.5} $\pm$ 0.17}                                                                     & \multicolumn{3}{c|}{72.6 $\pm$ 0.12}                                                                              & \multicolumn{3}{c|}{79.2 $\pm$ 0.25}                                                         \\ \cline{2-12}
\multicolumn{1}{|c|}{}                                                                                      & \multicolumn{2}{c|}{\textbf{Operand}}                                                                           & \multicolumn{3}{c|}{\textbf{86.2} $\pm$ 0.17}                                                                     & \multicolumn{3}{c|}{74.8 $\pm$ 0.19}                                                                              & \multicolumn{3}{c|}{71.1 $\pm$ 0.28}                                                         \\ \hline
\multicolumn{3}{|c|}{}                                                                                                                                                                                                        & \multicolumn{1}{c}{\textit{Prec.}} & \multicolumn{1}{c}{\textit{Rec.}} & \multicolumn{1}{c|}{\textit{nDCG}} & \multicolumn{1}{c}{\textit{Prec.}} & \multicolumn{1}{c}{\textit{Rec.}} & \multicolumn{1}{c|}{\textit{nDCG}} & \multicolumn{1}{c}{\textit{Prec.}} & \multicolumn{1}{c}{\textit{Rec.}} & \textit{nDCG} \\ \hline
\multicolumn{1}{|c|}{\multirow{6}{*}{\textbf{Similarity}}}                                                  & \multicolumn{1}{c|}{\multirow{3}{*}{\textbf{Strands}}}                                                 & top-10 & \multicolumn{1}{c}{\textbf{42.7}}  & \multicolumn{1}{c}{\textbf{44.0}} & \multicolumn{1}{c|}{\textbf{61.1}} & \multicolumn{1}{c}{42.4}           & \multicolumn{1}{c}{43.6}          & \multicolumn{1}{c|}{60.9}          & \multicolumn{1}{c}{35.7}           & \multicolumn{1}{c}{36.8}          & 53.9          \\
\multicolumn{1}{|c|}{}                                                                                      & \multicolumn{1}{c|}{}                                                                                  & top-25 & \multicolumn{1}{c}{\textbf{22.9}}  & \multicolumn{1}{c}{\textbf{53.5}} & \multicolumn{1}{c|}{\textbf{61.6}} & \multicolumn{1}{c}{22.7}           & \multicolumn{1}{c}{52.8}          & \multicolumn{1}{c|}{61.3}          & \multicolumn{1}{c}{18.9}           & \multicolumn{1}{c}{43.6}          & 53.5          \\
\multicolumn{1}{|c|}{}                                                                                      & \multicolumn{1}{c|}{}                                                                                  & top-40 & \multicolumn{1}{c}{\textbf{16.5}}  & \multicolumn{1}{c}{\textbf{57.9}} & \multicolumn{1}{c|}{\textbf{63.0}} & \multicolumn{1}{c}{16.1}           & \multicolumn{1}{c}{56.7}          & \multicolumn{1}{c|}{62.3}          & \multicolumn{1}{c}{13.4}           & \multicolumn{1}{c}{47.0}          & 54.3          \\ \cline{2-12}
\multicolumn{1}{|c|}{}                                                                                      & \multicolumn{1}{c|}{\multirow{3}{*}{\textbf{\begin{tabular}[c]{@{}c@{}}Basic \\ Blocks\end{tabular}}}} & top-5  & \multicolumn{1}{c}{\textbf{55.3}}  & \multicolumn{1}{c}{\textbf{25.4}} & \multicolumn{1}{c|}{61.2} & \multicolumn{1}{c}{54.9}           & \multicolumn{1}{c}{25.0}          & \multicolumn{1}{c|}{{\bf 61.3}}          & \multicolumn{1}{c}{52.7}           & \multicolumn{1}{c}{24.1}          & 59.2          \\
\multicolumn{1}{|c|}{}                                                                                      & \multicolumn{1}{c|}{}                                                                                  & top-10 & \multicolumn{1}{c}{\textbf{38.9}}  & \multicolumn{1}{c}{\textbf{34.5}} & \multicolumn{1}{c|}{\textbf{50.1}} & \multicolumn{1}{c}{38.3}           & \multicolumn{1}{c}{33.9}          & \multicolumn{1}{c|}{49.9}          & \multicolumn{1}{c}{35.8}           & \multicolumn{1}{c}{31.8}          & 47.4          \\
\multicolumn{1}{|c|}{}                                                                                      & \multicolumn{1}{c|}{}                                                                                  & top-25 & \multicolumn{1}{c}{\textbf{19.5}}  & \multicolumn{1}{c}{\textbf{42.5}} & \multicolumn{1}{c|}{\textbf{48.9}} & \multicolumn{1}{c}{19.5}           & \multicolumn{1}{c}{42.2}          & \multicolumn{1}{c|}{48.9}          & \multicolumn{1}{c}{18.2}           & \multicolumn{1}{c}{39.6}          & 46.4          \\ \hline
\end{tabular}%
\vspace{5pt}
\caption{Intrinsic evaluation results.}
\label{table:intrinsic_res}
\end{table*}

\begin{figure*}[t!]
\centering
\subfloat[Precision for the  top-$k$ answers with $k \leq 50$. \label{fig:precision} ]{\includegraphics[width=.33\textwidth]{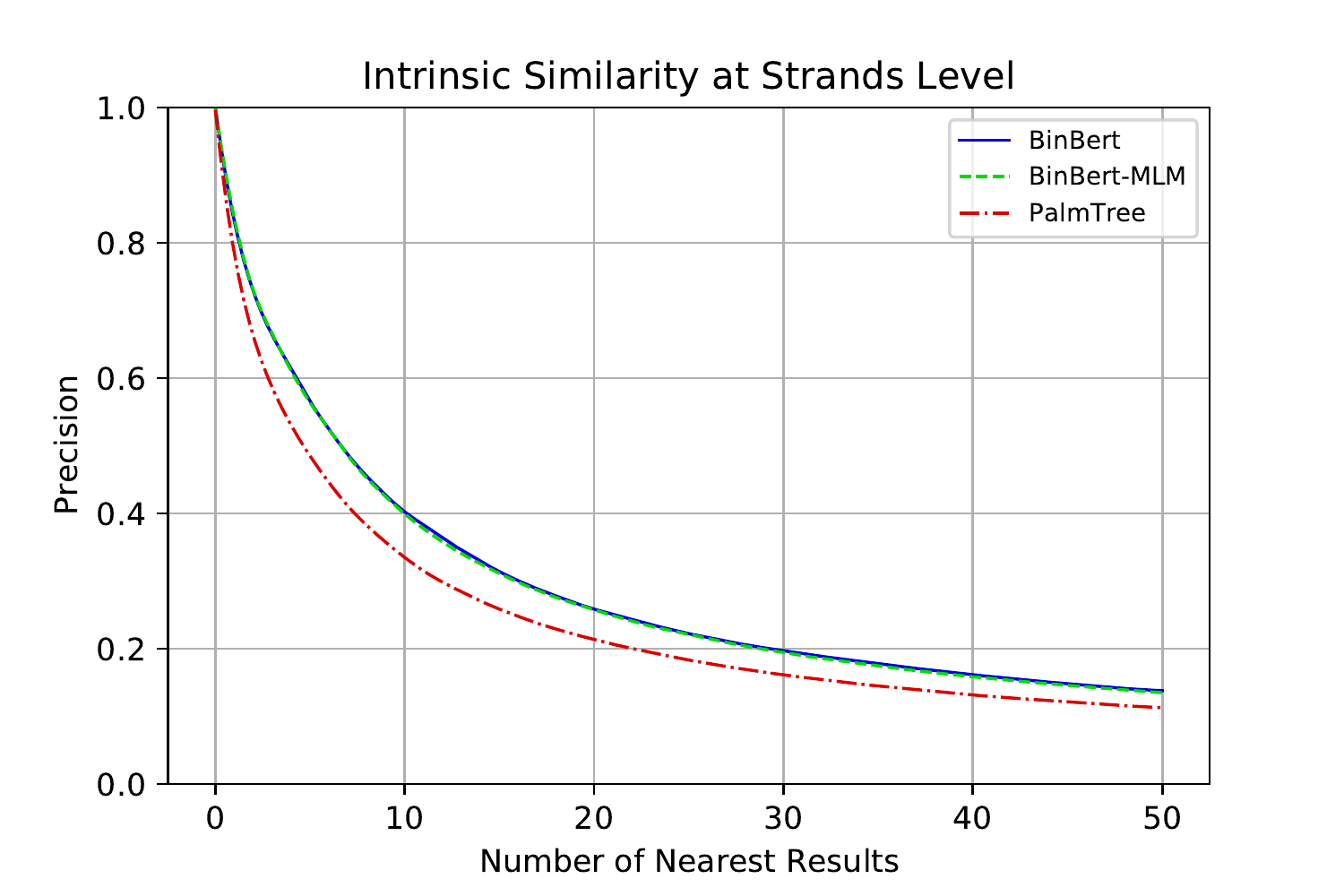}}\hfill
\subfloat[nDCG for the top-$k$ answers with $k \leq 50$. \label{fig:ndcg} ]{\includegraphics[width=.33\textwidth]{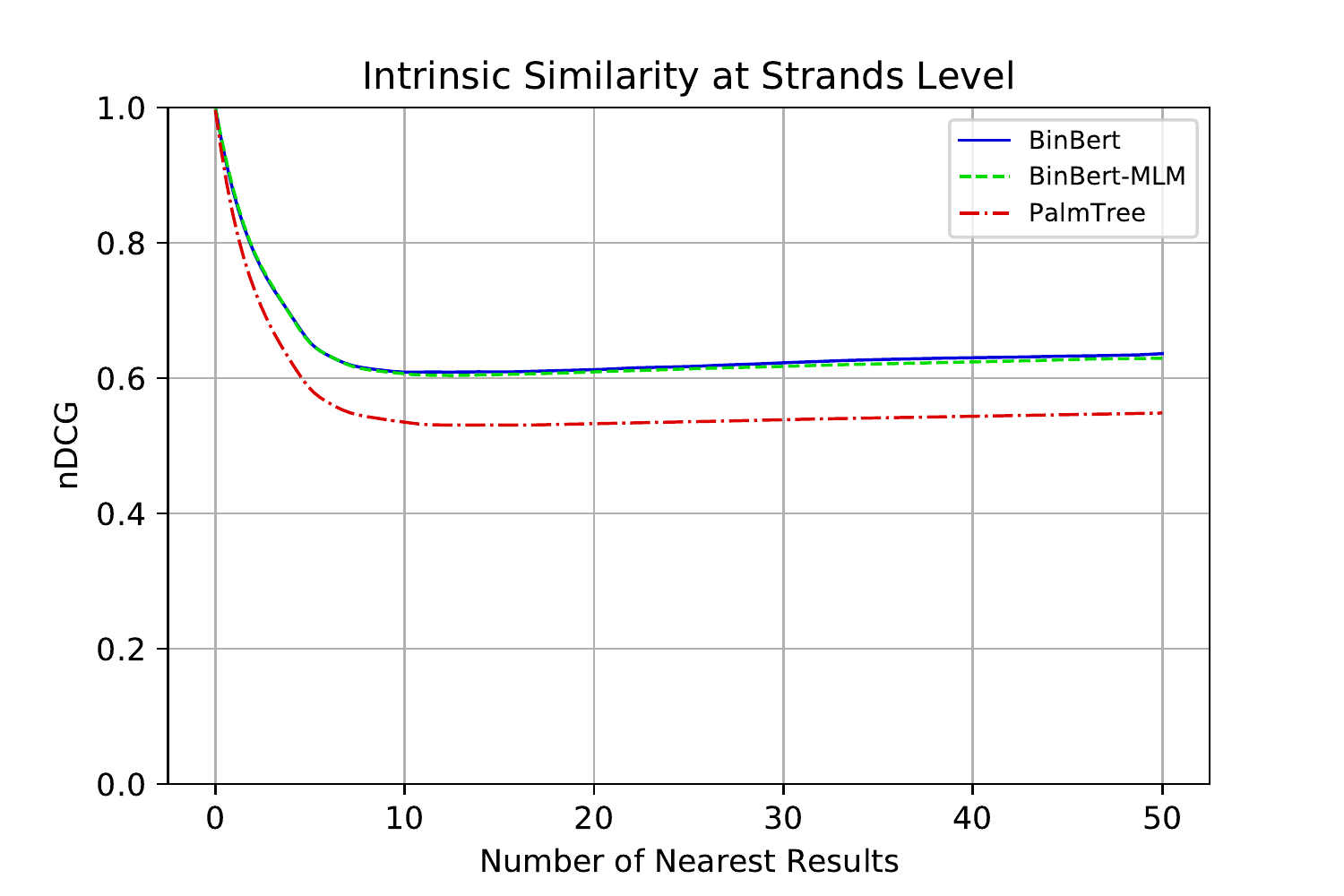}}\hfill
\subfloat[Recall for the  top-$k$ answers with $k \leq 50$. \label{fig:recall} ]{\includegraphics[width=.33\textwidth]{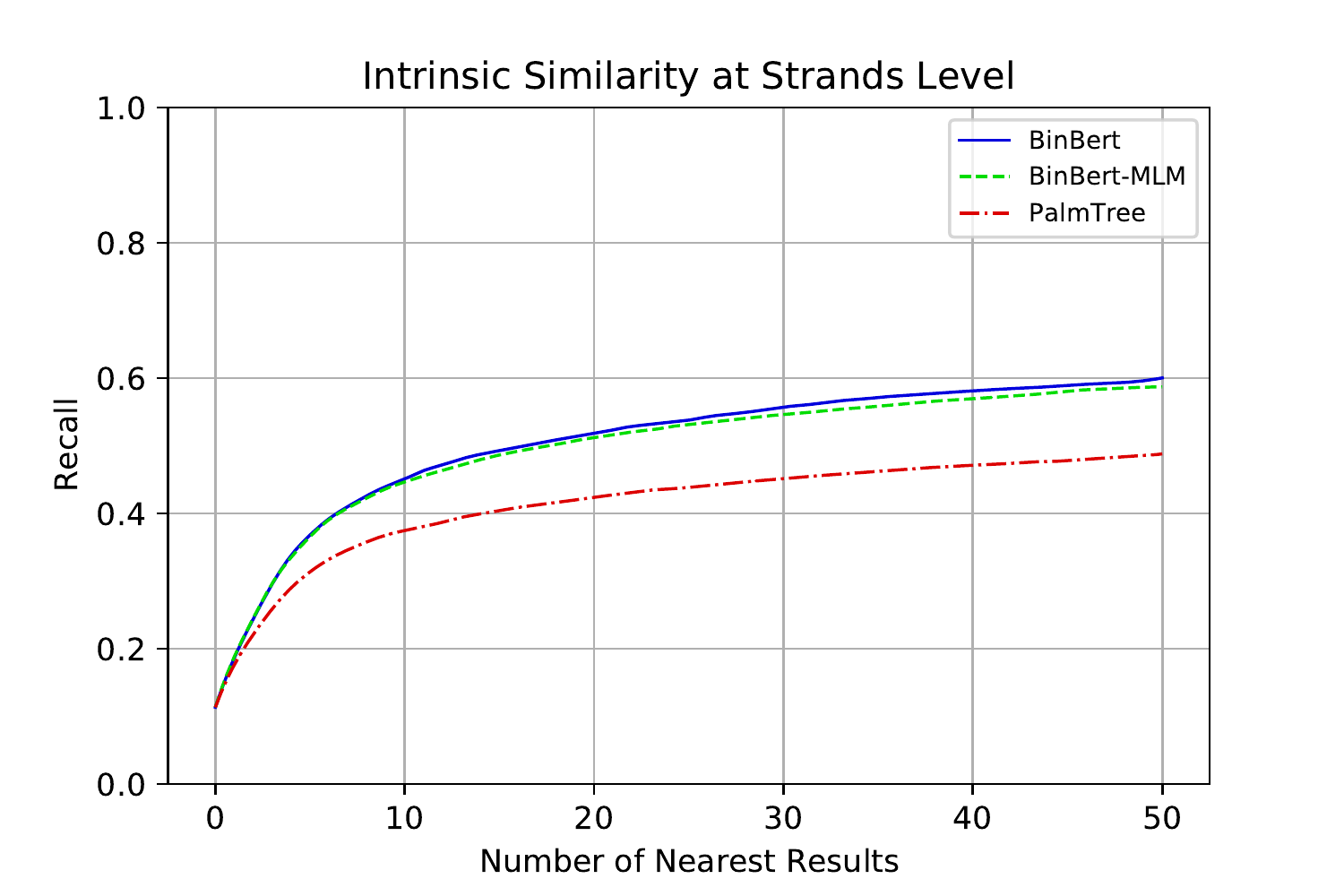}}\hfill
\caption{  Results for the {\bf intrinsic strand similarity} task. Database of 49079 strands, average on 833 queries. \label{fig:strandsearch} \color{black}}
\end{figure*}

\subsubsection{Opcode and Operand Outlier}
\label{sec:out_detection}
We used 43879 different instructions to create a dataset of 50000 sets of 5 instructions. These sets are created according to the task (either opcode or operand) as we have defined in Section \ref{subsec:outlier_detection}. 
We use opcode and operand categories analogous to \cite{palmtree2021li}: opcodes are categorized according to the \textit{x86 Assembly Language Reference Manual} \footnote{ \url{https://docs.oracle.com/cd/E26502_01/html/E28388/ennbz.html}}, while operands are categorized according to their type (e.g. two registers, one register and one memory access, etc.). Further details about these classes can be found in the Appendix \ref{appendix:op_class}.
To solve this task, we embed each instruction in the set and we evaluate if the embeddings are able to distinguish the outlier; this is done by computing the distance of each embedded instruction from the others and by predicting as outlier the most distant. An instruction embedding is computed by mean pooling the instruction tokens' hidden states in the second last layer of BinBert (we take the second last layer as it is less influenced by the pre-training task). 
The evaluation metric is the \textit{accuracy} of \cite{camacho-navigli-2016-find}:

\begin{equation}
Accuracy = \frac{\sum_{s \in \mathbb{S}} outlier(s)}{|\mathbb{S}|}
\label{eq:out_accuracy}
\end{equation}
where $\mathbb{S}$ is the dataset composed of instruction sets and $outlier(s)$ is equal to 1 if the outlier in the instruction set $s$ is detected and 0 otherwise. 

We compute the mean accuracy and standard deviation on 10 runs of the experiment on different datasets (each composed of 50k sets), the results are in Table \ref{table:intrinsic_res}. BinBert achieves the best performances (0.84 on opcodes/0.86 on operands), it shows a great improvement over BinBert-MLM (0.73 on opcodes/0.75 on operands) this confirms that symbolic expressions clearly enrich the semantic learned by the model for each instruction. It also outperforms PalmTree (0.79 on opcodes/0.71 on operands) by a wide margin; the performances of PalmTree are not dominated by a transfomer trained on MLM. We believe that this is so because PalmTree has been explicitly designed to be an instruction embedding solution, while BinBert-MLM has been trained on sequences. In Appendix \ref{appendix:op_class} we report a qualitative analysis with the clusters of opcodes learned by BinBert.

\subsubsection{Similarity at Strand Level} \label{sec:instrand}
We use a database $\mathbb{A}$ of 49079 strands, from this database we extract a database of 833 queries $\mathbb{Q}$. On average for each query we have 59 similar elements in $\mathbb{A}$. 
To solve the task we compute an embedding vector $\vec{q}$ for each query $q$ and an embedding vector $\vec{a}$ for each strand $a$ in the lookup database $\mathbb{A}$. This is done by averaging all the instruction tokens' hidden states in the second last layer of BinBert. For each query vector $\vec{q}$ we compute the cosine similarity with all $\vec{a} \in \mathbb{A}$; we return the ordered list of the top-k similar elements $R_{q}=(r_1, \ldots, r_k)$. Using $R_{q}$ we compute: precision, number of true similars in  $R_{q}$  over $k$; recall, number of true similar in $R_{q}$ over $\#sim(q)$, that is the number of items similar to $q$ in $\mathbb{A}$; and nDCG. The nDCG is a measure used in information retrieval. It is defined as:

\begin{equation}
nDCG= \frac{\sum_{i=1}^k \frac{sim(r_i, q)}{log(1+i)}}{\sum_{i=1}^{\#sim(q)} \frac{1}{log(1+i)}}
\end{equation}

Where $sim(r_i,q)$ is $1$ if $q$ is similar to $r_i$ and 0 otherwise. The quantity at the denominator is the scoring of a perfect answer, and the number at the numerator is the scoring of our system. The nDCG is between $0$ and $1$, and it takes into account the ordering of the items in $R_{q}$, giving better scores when similar items are ordered first.  
As an example let us suppose we have two results for the same query: $(1,1,0,0)$ and $(0,0,1,1)$ (where $1$ means that the corresponding index in the result list is occupied by a similar item and $0$ otherwise). These results have the same precision (i.e., $\frac{1}{2}$), but nDCG scores the first better. 
 We average the per-query precision, recall, and nDCG to obtain the final metrics. Results are shown in Table \ref{table:intrinsic_res} and in Figure \ref{fig:strandsearch}.

We can see that BinBert achieves the best performance on precision, recall and nDCG.
%For $k \in \{10,25,40\}$ the precision values of BinBert are  $[42.7, 22.9, 16.5]$ this is a marked gap with respect to PalmTree ($[35.7,18.9,13.4]$), the same behaviour can be observed for recall and nDCG. \GADL{calcolare il range di aumento di performance}
%It is interesting to note that the gap is not as wide with  in Bert-MLM. The execution-awareness still gives the edge to BinBert on all measures and for all values of $k$, \GADL{abbiamo una spiegazione?}.

\subsubsection{Similarity at Block Level} \label{sec:inblock}
To solve this task we used a database $\mathbb{A}$ of 6460 basic blocks where 505 are queries $\mathbb{Q}$. On average each query has 10 similar elements in $\mathbb{A}$. The test procedure and the metrics are the same of the strand similarity case except for the basic block embedding computation. A basic block is first decomposed into strands and then its embedding is obtained as the average of its strand embeddings. Results are shown in Table \ref{table:intrinsic_res} (the figures for the values of $k \in [0,30]$ are reported in the Appendix).
We can see the same general behavior observed with the strand similarity, with BinBert being the top performer. But there is less gap with respect to PalmTree. We believe that this is due to the fact that PalmTree is not trained directly on strands, and so the strand similarity task is harder for PalmTree than block similarity. It is interesting to note that the reverse effect is not observable for BinBert, the performances on blocks are slightly better than the ones on strands.

\begin{mdframed}[backgroundcolor=mPurple!8]

 \begin{description}[style=unboxed,leftmargin=0cm]
Our evaluation shows that the execution-aware transformer BinBert is state-of-the-art for all the intrinsic tasks of our benchmark. Execution-awareness improves performances on all tasks.  The benefit is greater on tasks that measure the understanding of instructions semantic.
\end{description}

\end{mdframed}

\subsection{Extrinsic Tasks at Strand and Block Level}
Results for extrinsic tasks are shown in Table \ref{table:extrinsic_res}.

\begin{figure}[t!]
\center
    \includegraphics[scale=0.65]{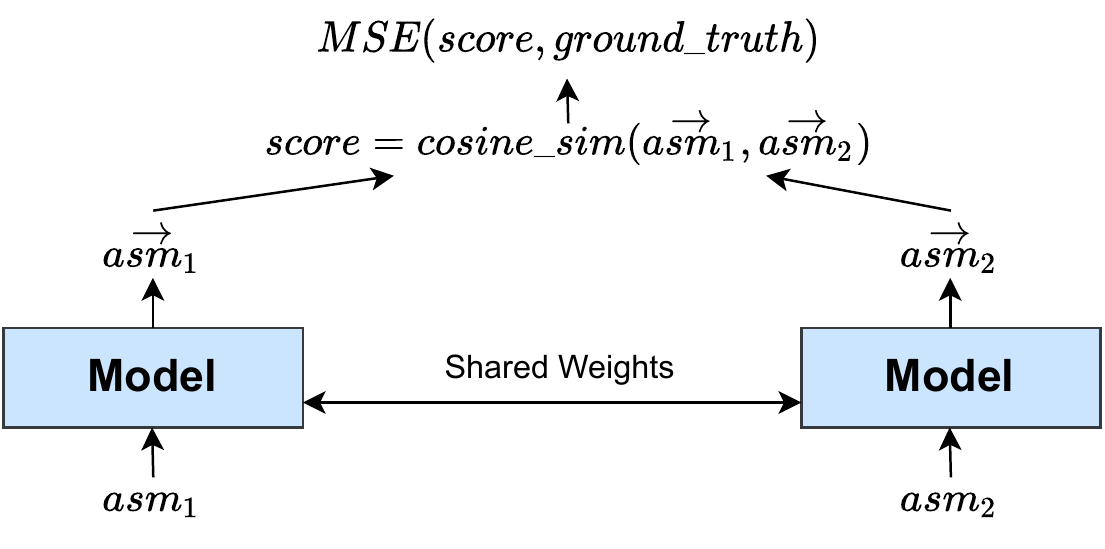}
    \vspace{-0.4cm}
    \caption{Scheme for the siamese architecture: $asm_i$ can be either a strand, basic block or a function and Model is either BinBert, BinBert-ML or PalmTree.
\label{fig:siamese}}
\end{figure}
 
\subsubsection{Strands and Blocks Similarity}
\label{sec:sb_similarity}

\paragraph{Fine-tune Process and Dataset}
We fine-tuned BinBert, BinBert-MLM, and PalmTree on the task of recognizing if a pair of entities is similar or dissimilar. Considering strands, we construct a dataset of  44978 pairs of strands, 50\% similar and 50\% dissimilar.
We split it into train and validation, resulting in 39979 strands pairs for the training set and 4999 for the validation. For blocks, we construct an analogous dataset with 24001 basic blocks pairs for the training set and 3001 pairs for the validation.

We fine-tune using a siamese architecture \cite{siamese,safe19mass, gemini}  (see Figure \ref{fig:siamese}). In this architecture two instances of the embedding network are used, each instance produces the embedding of the corresponding entity in a pair. The resulting embeddings are compared with the cosine similarity to produce a score. The siamese network is trained with the \textit{Mean Squared Error Loss} (\textit{MSE}) which minimizes the distance between the similarity score produced by the network and the ground truth label that can be in $\{+1,-1\}$ (i.e. two strands/basic blocks are similar or not). Intuitively, this training process instructs the embedding network to produce embeddings that are close if they are from a similar pair and distant if not. 
For BinBert-FT and BinBert-MLM-FT the embedding that enters the cosine similarity is the average of all tokens of the last layer (excluded the padding).
 
Since PalmTree is an instruction embedding model, an additional architecture is needed to transform it into a model that embeds sequences. In particular, we use a bidirectional LSTM, where each cell takes as input the instruction embedding produced by PalmTree. Finally, we compute an embedding by averaging all the hidden states of the LSTM.
We test the fine-tuned models on the same test sets used in the corresponding intrinsic tasks (see Sec.\ref{sec:it}).  We fine-tune each model for 20 epochs, selecting the epoch with the best AUC on validation.   

\paragraph{Results}

\begin{figure*}[htbp]
\centering
\subfloat[nDCG for the top-$k$ answers with $k \leq 50$.]{\includegraphics[width=0.45\textwidth]{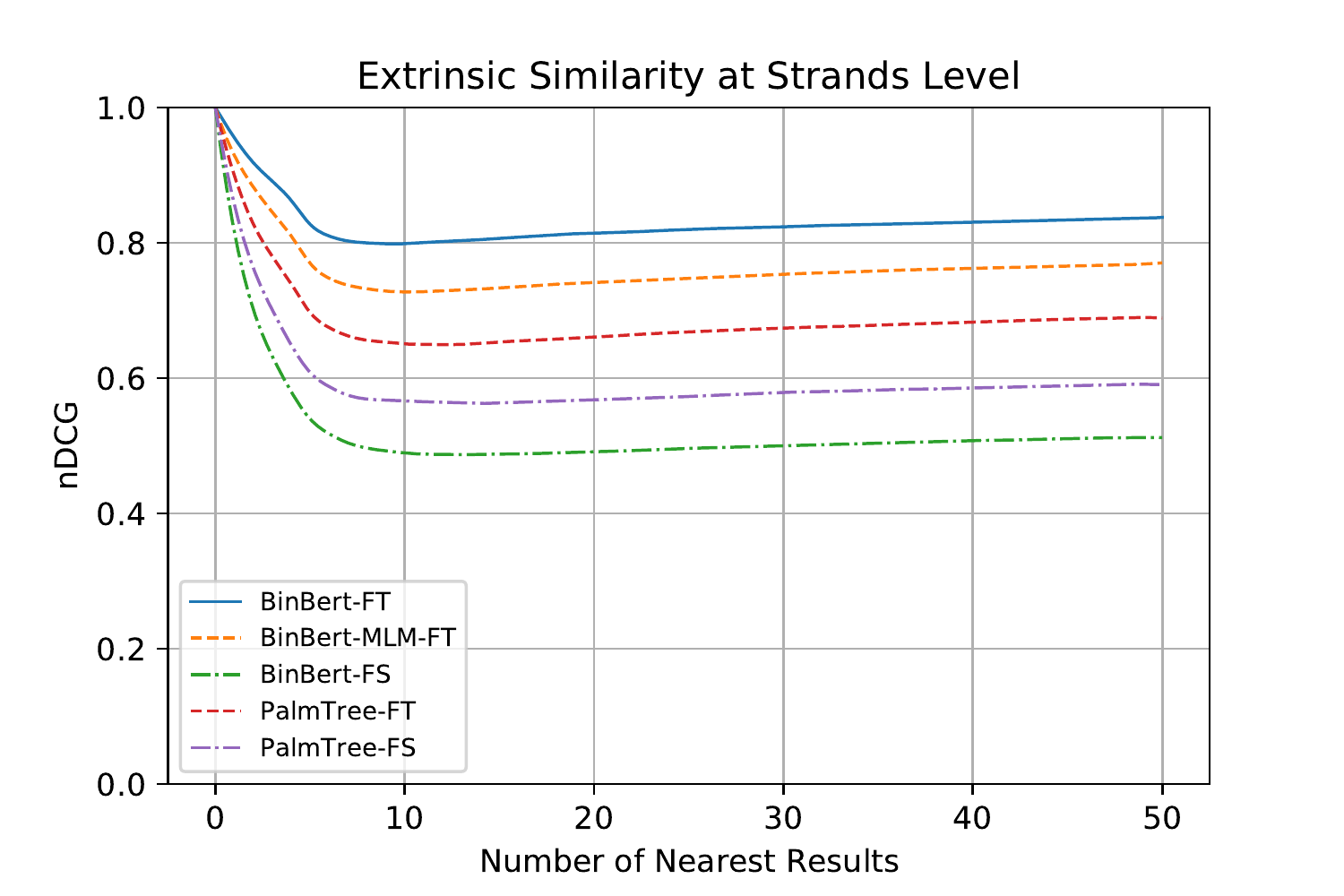}}\hfill
\subfloat[Recall for the top-$k$ answers with $k \leq 50$.] {\includegraphics[width=0.45\textwidth]{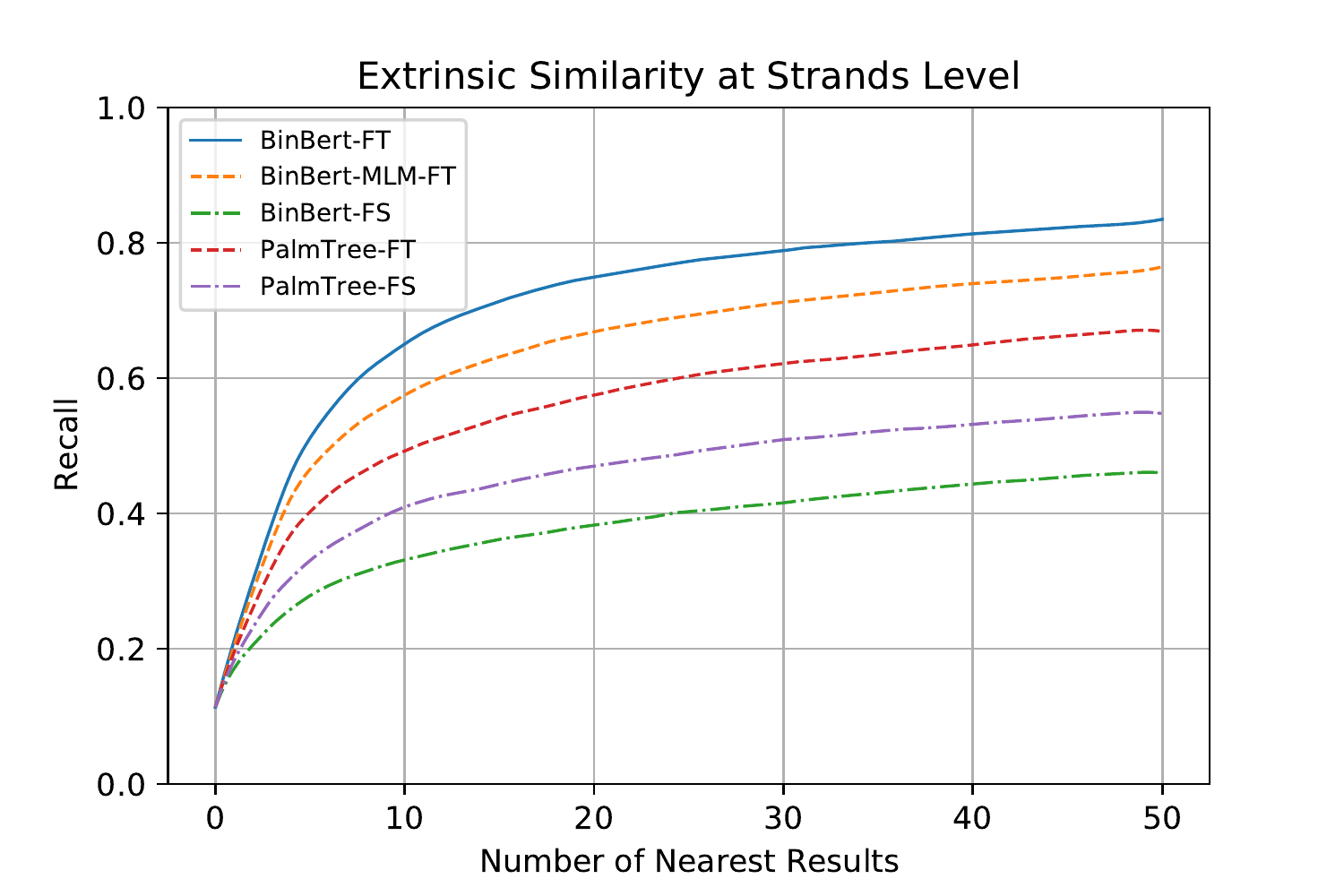}}\hfill
\caption{Results for the {\bf extrinsic strand similarity} task. Database of 49079 strands, average on 833 queries.} \label{fig:exstrand}
\end{figure*}

Results for strands and basic blocks similarities are in Table \ref{table:extrinsic_res}. In Figures \ref{fig:exstrand} there are the results of strand similarity for $k \in [0,50]$, the figures for block similarity are in the Appendix.

Also in this case BinBert achieves the best performances. The gap between BinBert-FT and the other models is much wider than in the intrinsic case. A possibility is that BinBert learns a wider semantic during pre-training, solving more efficiently the similarity task after fine-tuning. This explains the gap between BinBert-FT and BinBert-MLM-FT. 

The great impact of pre-training can be appreciated by looking at the performance of BinBert-FT and BinBert-FS. BinBert-FS has been only trained on the fine-tune dataset so it cannot leverage a learned semantic, the fine-tune dataset has not enough data to make up for this disadvantage and to train a big transformer model. The difference in model size between PalmTree-FS and BinBert-FS is the reason why PalmTree-FS performs better than BinBert-FS, a smaller model can be trained with less data. 

The impact of pre-training can also be seen on PalmTree, PalmTree-FT beats PalmTree-FS. In this case, the difference is less marked than the one between BinBert-FT and BinBert-FS, the reasons probably are: PalmTree is a smaller model (and so it can be trained and reach a plateau with less data), and being an instruction embedding model cannot learn during pre-training the many complex interactions between instructions that can be useful for a sequence similarity task.

\subsubsection{Compiler Provenance}
\label{par:compiler_provenance}
As in other works on compiler provenance \cite{InvestigatingGE19mass}, we train and test the networks on the task of {\em Compiler Classification}, that is detecting the compiler family that has generated a sample, and {\em Optimization Classification}, detecting the optimization level used to generate a sample. 

\paragraph{Fine-tune Process and Dataset}
The compiler provenance dataset is made up of samples in which, depending on the granularity, a strand or a basic block is associated with two labels: the compiler family with its corresponding version and the optimization used. The strand compiler provenance dataset has 36828 samples, split into 29462 samples for the training set and 3683 samples for both the validation and test sets. The basic blocks compiler provenance dataset has 81918 samples, split with the same ratio as before, resulting in 65534 training samples and 8192 validation and test samples.

We fine-tune BinBert and BinBert-MLM on both compiler and optimization classification by adding a linear layer followed by the softmax function on top of the last layer hidden state corresponding to the \asm{[CLS]} token. For PalmTree we use an LSTM over the instruction embedding tokens generated; to obtain a classifier we attach a linear layer with the softmax on top of the last hidden states of the RNN.  We fine-tune each model for 20 epochs, selecting the epoch with the best classification accuracy on validation.   

\paragraph{Results}
Results for both compiler and optimization classification are shown in Table \ref{table:extrinsic_res}. The compiler classification task is the only task where BinBert-FT is slightly worse than BinBert-MLM-FT, we believe that this is due to the fact that recognising a compiler signature is a syntactic task. 
PalmTree-FT has the worse performance among all fine-tuned models. Again we can see that pre-training is important as all the from-scratch models perform consistently worse than their fine-tuned counterpart.  Results are similar for the optimization classification task.

\subsubsection{Strand Recovery}
\paragraph{Fine-tune Process and Dataset}
The dataset is made of 9267 basic blocks, each block contains at least 5 disjoint strands (i.e. the strands do not overlap on instructions). We split the dataset into 7412 training samples and 927 samples for both validation and test. We model strand recovery as instruction classification; given the instructions of a basic block and the final instruction of one of its strands, we aim at classifying the other instructions as either belonging to the same strand as the marked instruction or not. To mark an instruction we surround it with a special token.

We fine-tune BinBert and BinBert-MLM by attaching a classification head on top of the last layer hidden states of the first token of each instruction; the network will output 1 if an instruction is part of the strand to be recovered and 0 otherwise. As for previous tasks, we use an LSTM on top of PalmTree and we put a classification head on the hidden states of the first token of each instruction. We fine-tune each model for 20 epochs, selecting the epoch with the best classification accuracy on validation.

\paragraph{Results}
We reported precision, recall, and F1-score of the positive class in Table \ref{table:extrinsic_res}. We can see that the best performing model is BinBert-FT, it markedly surpasses PalmTree-FT on all the metrics considered. BinBert-MLM-FT is the second best model, it achieves the same precision as BinBert-FT but smaller recall. We believe the reason to be the execution-awareness of BinBert-FT that allows the model to recover more instructions in a strand. %A qualitative analysis showing the change of internal attention weights of BinBert during fine-tuning is in Appendix \ref{app:attention}. 

% !TEX root =  main.tex

\subsubsection{Strand Execution}
\paragraph{Fine-tune Process and Dataset}
We created a dataset for the strand execution task by taking strand-symbolic execution pairs, assigning concrete values to input variables, and evaluating its concrete output. In particular, we randomly assign values between 0 and 100 to input variables and we take only strands whose output is not greater than 200. Our dataset only contains strands computing the value of a register or a predicate of a conditional branch. The dataset is composed of 40000 training samples, and 5000 validation and test samples (total 50k). Each sample is made up of strand instructions followed by concrete assignments of input variables and the query output variables (only in the case of register outputs); the label for such a sample is the concrete value of the output variable. For instance, consider the strands \asm{mov eax, dword ptr [rbp - 180] sub eax, 1} and the value 9 assigned to \asm{dword ptr [rbp - 180]}. The corresponding sample will be \asm{mov eax, dword ptr [rbp - 180] sub eax, 1 [SEP] dword ptr [rbp - 180] = 9 [SEP] rax}. The network has to predict the value for register \asm{rax}, in this case 8.

We model this problem as a sequence classification task, thus we used the same architectures used for the compiler provenance tasks (see Section \ref{par:compiler_provenance}).

\paragraph{Results}
We reported the accuracy obtained by all the models in Table \ref{table:extrinsic_res}. BinBert-FT achieves the best performances. PalmTree-FT performs poorly on this task, we believe that this is due to the fact that PalmTree is a pure instruction embedding model so even when fine-tuned it cannot transfer to the upward neural architecture $A$ sequence-related knowledge. On the contrary, both BinBert-FT and BinBert-MLM-FT are trained on sequences; however, BinBert-FT has the edge thanks to its execution-awareness. In this case, the pre-training has a great impact as we can see by the poor performance of BinBert-FS. 

\subsection{Extrinsic Tasks at Function Level}

\subsubsection{Similarity}
\paragraph{Fine-tune Process and Dataset}
The function similarity dataset has exactly the same format as the similarity tasks at strand and block-level (see Section \ref{sec:sb_similarity}). In particular, the training dataset contains 18805 function pairs, while the validation set contains 1675 pairs. The test set is composed of 3627 functions, where 267 are queries. On average each query has 14 similar entities in the dataset.

On top of the embedding models, we use the same architectures that we employed to solve the other similarity tasks. As in  \cite{safe19mass} we truncate all functions to the first 150 instructions.

We compare our solution with Asm2vec \cite{asm2vec} and Safe \cite{safe19mass}, which are solutions specifically crafted for the binary similarity problem. Asm2vec uses PV-DM \cite{pvdm} to simultaneously learn instruction and function embeddings, while Safe uses word2vec \cite{w2v} to create instruction embeddings to be fed to a self-attentive neural network \cite{sa17bengio} for learning the final embeddings for functions.

\begin{figure*}[htbp]
\centering
\subfloat[nDCG for the top-$k$ answers with $k \leq 30$.]{\includegraphics[width=0.45\textwidth]{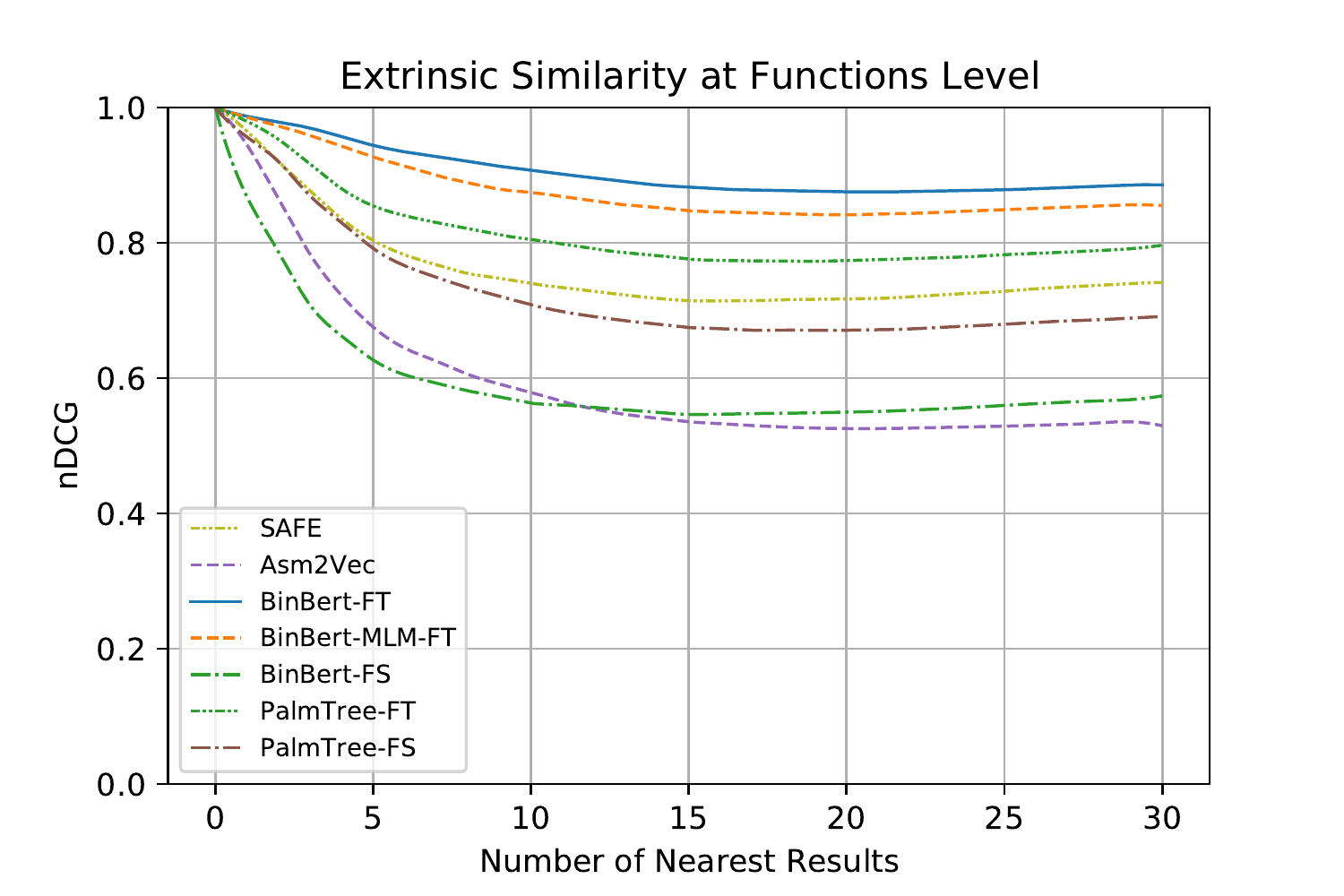}}\hfill
\subfloat[Recall for the top-$k$ answers with $k \leq 30$.] {\includegraphics[width=0.45\textwidth]{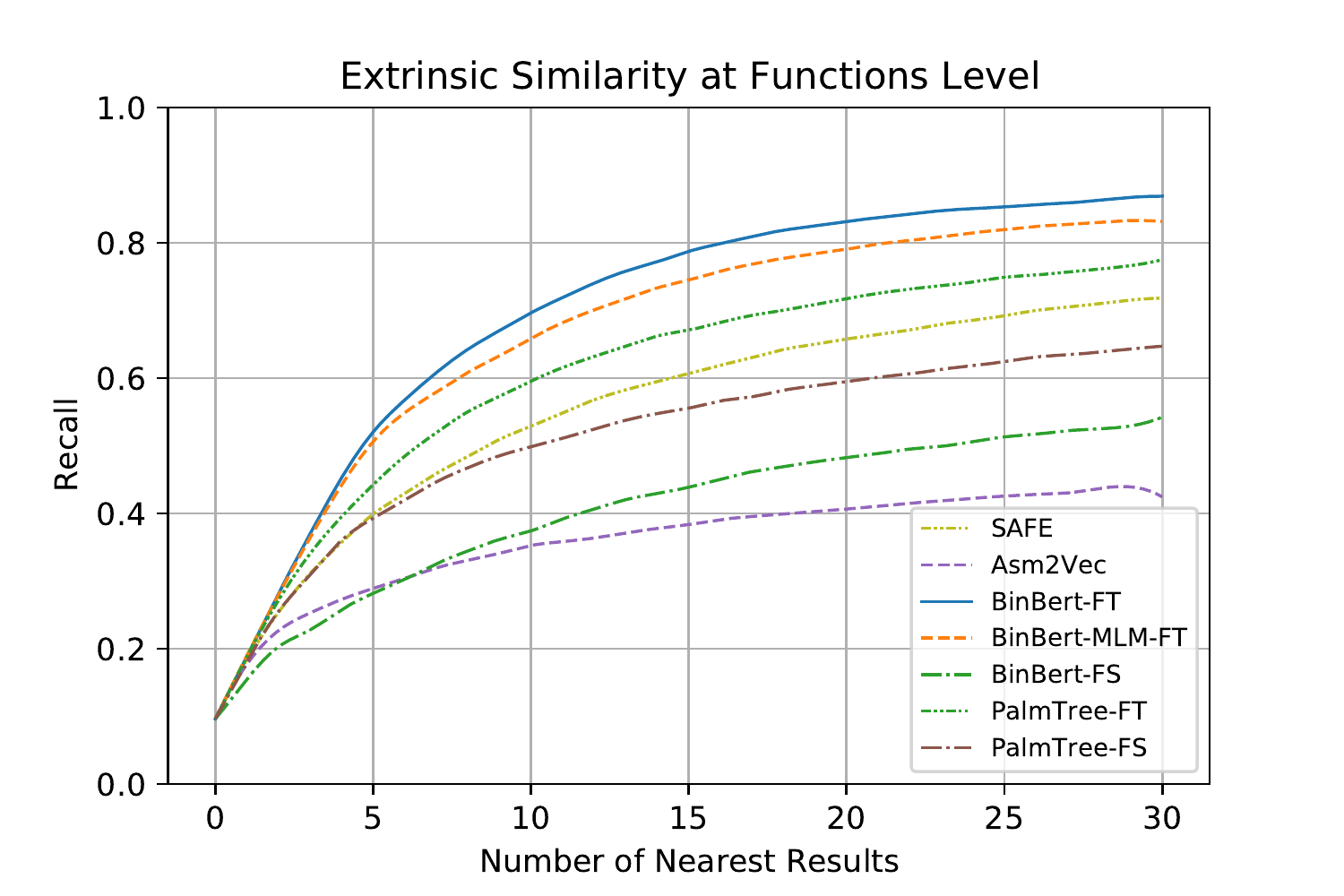}}\hfill
\caption{Results for the {\bf function similarity} task.} \label{fig:exfunc}
\end{figure*}

\paragraph{Results}
Results are in Figure \ref{fig:exfunc}. BinBert achieves the best performances, beating also Asm2vec and Safe which are specific for the function similarity problem. PalmTree performs worse than BinBert, confirming our initial intuition about the drawbacks of the lack of context and the isolated instruction embeddings.  
Asm2vec is the worst performing model and is beaten by PalmTree (this aligns with the results in \cite{palmtree2021li}). Safe has lower performances than PalmTree. This suggests that even if they are both based on fixed instructions embeddings, a transformer encoder is more capable of capturing an instruction semantic than word2vec.
Also in this task, BinBert-FS and PalmTree-FS are below their corresponding fine-tuned versions (BinBert-FT and PalmTree-FT) thus highlighting the importance of pre-training. Finally, the impact of execution awareness can be seen in the advantage of BinBert-FT on BinBert-MLM-FT. 

\subsubsection{Compiler Provenance}
\paragraph{Fine-tune Process and Dataset}
The compiler provenance dataset has the same structure as the strand and block compiler provenance tasks. It contains 39017 functions in the training set and 4877 functions in the validation and test set. 

We used the same architectures that we employed to solve the other compiler provenance tasks. We truncate all the functions to the first 150 instructions (as done in \cite{safe19mass}).

\paragraph{Results}
We reported the accuracy for different models in Table \ref{table:extrinsic_res}. Results are similar to the compiler provenance task at strand and block level, thus the same considerations hold. In Appendix \ref{appendix:confusionmatrices} we report the confusion matrices for this experiment. 

\begin{mdframed}[backgroundcolor=mPurple!8]

 \begin{description}[style=unboxed,leftmargin=0cm]
Results from the experimental evaluation confirm that BinBert is the current state-of-the art for assembly code models. It shows improvement over PalmTree and deep learning solutions specifically tailored for a certain task. The execution awareness has a marked impact on semantic tasks. Interestingly, on some syntactic tasks, execution awareness does not increase the performance of the model. 
Our evaluation highlights the great impact of pre-training on downstream tasks with small size datasets. 
\end{description}

\end{mdframed}

\begin{table*}[t!]
\resizebox{\textwidth}{!}{%
\begin{tabular}{|ccc|ccccccccccccccc|}
\hline
\multicolumn{3}{|c|}{\multirow{3}{*}{\textbf{Tasks}}}                                                                                                                                                                                   & \multicolumn{15}{c|}{\textbf{Models}}                                                                                                                                                                                                                                                                                                                                                                                                                                                                                                                    \\ \cline{4-18} 
\multicolumn{3}{|c|}{}                                                                                                                                                                                                                  & \multicolumn{3}{c|}{\textbf{BinBert-FT}}                                                                         & \multicolumn{3}{c|}{\textbf{BinBert-MLM-FT}}                                                                     & \multicolumn{3}{c|}{\textbf{BinBert-FS}}                                                                      & \multicolumn{3}{c|}{\textbf{PalmTree-FT}}                                                                        & \multicolumn{3}{c|}{\textbf{PalmTree-FS}}                                                \\ \cline{4-18} 
\multicolumn{3}{|c|}{}                                                                                                                                                                                                                  & \multicolumn{1}{c}{\textit{Prec.}} & \multicolumn{1}{c}{\textit{Rec.}} & \multicolumn{1}{c|}{\textit{nDCG}} & \multicolumn{1}{c}{\textit{Prec.}} & \multicolumn{1}{c}{\textit{Rec.}} & \multicolumn{1}{c|}{\textit{nDCG}} & \multicolumn{1}{c}{\textit{Prec.}} & \multicolumn{1}{c}{\textit{Rec.}} & \multicolumn{1}{c|}{\textit{nDCG}} & \multicolumn{1}{c}{\textit{Prec.}} & \multicolumn{1}{c}{\textit{Rec.}} & \multicolumn{1}{c|}{\textit{nDCG}} & \multicolumn{1}{c}{\textit{Prec.}} & \multicolumn{1}{c}{\textit{Rec.}} & \textit{nDCG} \\ \hline
\multicolumn{1}{|c|}{\multirow{9}{*}{\textbf{Similarity}}}                                                            & \multicolumn{1}{c|}{\multirow{3}{*}{\textbf{Strands}}}                                                 & top-10 & \multicolumn{1}{c}{\textbf{60.9}}  & \multicolumn{1}{c}{\textbf{63.1}} & \multicolumn{1}{c|}{\textbf{79.9}} & \multicolumn{1}{c}{53.3}           & \multicolumn{1}{c}{55.9}          & \multicolumn{1}{c|}{72.9}          & \multicolumn{1}{c}{31.3}           & \multicolumn{1}{c}{32.4}          & \multicolumn{1}{c|}{49.2}          & \multicolumn{1}{c}{46.4}           & \multicolumn{1}{c}{48.0}          & \multicolumn{1}{c|}{65.3}          & \multicolumn{1}{c}{38.4}           & \multicolumn{1}{c}{39.8}          & 56.8          \\ 
\multicolumn{1}{|c|}{}                                                                                                & \multicolumn{1}{c|}{}                                                                                  & top-25 & \multicolumn{1}{c}{\textbf{33.3}}  & \multicolumn{1}{c}{\textbf{76.8}} & \multicolumn{1}{c|}{\textbf{81.9}} & \multicolumn{1}{c}{29.2}           & \multicolumn{1}{c}{68.8}          & \multicolumn{1}{c|}{74.7}          & \multicolumn{1}{c}{16.9}           & \multicolumn{1}{c}{40.0}          & \multicolumn{1}{c|}{49.6}          & \multicolumn{1}{c}{25.4}           & \multicolumn{1}{c}{59.8}          & \multicolumn{1}{c|}{66.7}          & \multicolumn{1}{c}{20.6}           & \multicolumn{1}{c}{48.5}          & 57.1          \\  
\multicolumn{1}{|c|}{}                                                                                                & \multicolumn{1}{c|}{}                                                                                  & top-40 & \multicolumn{1}{c}{\textbf{23.3}}  & \multicolumn{1}{c}{\textbf{81.1}} & \multicolumn{1}{c|}{\textbf{83.0}} & \multicolumn{1}{c}{20.7}           & \multicolumn{1}{c}{73.8}          & \multicolumn{1}{c|}{76.2}          & \multicolumn{1}{c}{12.2}           & \multicolumn{1}{c}{44.1}          & \multicolumn{1}{c|}{50.7}          & \multicolumn{1}{c}{18.1}           & \multicolumn{1}{c}{64.6}          & \multicolumn{1}{c|}{68.2}          & \multicolumn{1}{c}{14.8}           & \multicolumn{1}{c}{52.9}          & 68.6          \\ \cline{2-18} 
\multicolumn{1}{|c|}{}                                                                                                & \multicolumn{1}{c|}{\multirow{3}{*}{\textbf{\begin{tabular}[c]{@{}c@{}}Basic \\ Blocks\end{tabular}}}} & top-5  & \multicolumn{1}{c}{\textbf{60.1}}  & \multicolumn{1}{c}{\textbf{27.9}} & \multicolumn{1}{c|}{\textbf{65.7}} & \multicolumn{1}{c}{57.8}           & \multicolumn{1}{c}{26.5}          & \multicolumn{1}{c|}{63.3}          & \multicolumn{1}{c}{37.9}           & \multicolumn{1}{c}{17.4}          & \multicolumn{1}{c|}{46.6}          & \multicolumn{1}{c}{56.6}           & \multicolumn{1}{c}{26.0}          & \multicolumn{1}{c|}{62.5}          & \multicolumn{1}{c}{50.4}           & \multicolumn{1}{c}{23.0}          & 57.3          \\ 
\multicolumn{1}{|c|}{}                                                                                                & \multicolumn{1}{c|}{}                                                                                  & top-10 & \multicolumn{1}{c}{\textbf{44.6}}  & \multicolumn{1}{c}{\textbf{39.9}} & \multicolumn{1}{c|}{\textbf{55.5}} & \multicolumn{1}{c}{42.5}           & \multicolumn{1}{c}{37.9}          & \multicolumn{1}{c|}{53.4}          & \multicolumn{1}{c}{24.6}           & \multicolumn{1}{c}{22.0}          & \multicolumn{1}{c|}{36.0}          & \multicolumn{1}{c}{39.6}           & \multicolumn{1}{c}{35.4}          & \multicolumn{1}{c|}{51.1}          & \multicolumn{1}{c}{33.7}           & \multicolumn{1}{c}{30.0}          & 54.5          \\ 
\multicolumn{1}{|c|}{}                                                                                                & \multicolumn{1}{c|}{}                                                                                  & top-25 & \multicolumn{1}{c}{\textbf{23.7}}  & \multicolumn{1}{c}{\textbf{51.8}} & \multicolumn{1}{c|}{\textbf{55.9}} & \multicolumn{1}{c}{22.4}           & \multicolumn{1}{c}{48.9}          & \multicolumn{1}{c|}{53.5}          & \multicolumn{1}{c}{13.1}           & \multicolumn{1}{c}{29.0}          & \multicolumn{1}{c|}{35.9}          & \multicolumn{1}{c}{20.8}           & \multicolumn{1}{c}{45.2}          & \multicolumn{1}{c|}{51.0}          & \multicolumn{1}{c}{17.3}           & \multicolumn{1}{c}{37.7}          & 44.6          \\ \cline{2-18} 
\multicolumn{1}{|c|}{}                                                                                                & \multicolumn{1}{c|}{\multirow{3}{*}{\textbf{Functions}}}                                               & top-5  & \multicolumn{1}{c}{\textbf{94.3}}  & \multicolumn{1}{c}{\textbf{45.3}} & \multicolumn{1}{c|}{\textbf{95.7}} & \multicolumn{1}{c}{92.4}           & \multicolumn{1}{c}{44.2}          & \multicolumn{1}{c|}{94.3}          & \multicolumn{1}{c}{57.2}           & \multicolumn{1}{c}{25.8}          & \multicolumn{1}{c|}{66.2}          & \multicolumn{1}{c}{83.7}           & \multicolumn{1}{c}{39.5}          & \multicolumn{1}{c|}{88.0}          & \multicolumn{1}{c}{77.5}           & \multicolumn{1}{c}{36.0}          & 83.0          \\  
\multicolumn{1}{|c|}{}                                                                                                & \multicolumn{1}{c|}{}                                                                                  & top-10 & \multicolumn{1}{c}{\textbf{76.7}}  & \multicolumn{1}{c}{\textbf{67.0}} & \multicolumn{1}{c|}{\textbf{91.3}} & \multicolumn{1}{c}{72.5}           & \multicolumn{1}{c}{63.3}          & \multicolumn{1}{c|}{87.9}          & \multicolumn{1}{c}{41.7}           & \multicolumn{1}{c}{36.1}          & \multicolumn{1}{c|}{57.2}          & \multicolumn{1}{c}{65.3}           & \multicolumn{1}{c}{57.3}          & \multicolumn{1}{c|}{81.2}          & \multicolumn{1}{c}{55.2}           & \multicolumn{1}{c}{48.5}          & 72.2          \\ 
\multicolumn{1}{|c|}{}                                                                                                & \multicolumn{1}{c|}{}                                                                                  & top-25 & \multicolumn{1}{c}{\textbf{43.2}}  & \multicolumn{1}{c}{\textbf{85.1}} & \multicolumn{1}{c|}{\textbf{87.8}} & \multicolumn{1}{c}{41.2}           & \multicolumn{1}{c}{81.5}          & \multicolumn{1}{c|}{84.7}          & \multicolumn{1}{c}{24.2}           & \multicolumn{1}{c}{50.6}          & \multicolumn{1}{c|}{55.7}          & \multicolumn{1}{c}{36.7}           & \multicolumn{1}{c}{74.2}          & \multicolumn{1}{c|}{78.0}          & \multicolumn{1}{c}{29.8}           & \multicolumn{1}{c}{61.8}          & 67.7          \\ \hline
\multicolumn{3}{|c|}{}                                                                                                                                                                                                                  & \multicolumn{3}{c|}{\textit{Accuracy}}                                                                        & \multicolumn{3}{c|}{\textit{Accuracy}}                                                                        & \multicolumn{3}{c|}{\textit{Accuracy}}                                                                        & \multicolumn{3}{c|}{\textit{Accuracy}}                                                                        & \multicolumn{3}{c|}{\textit{Accuracy}}                                                   \\ \hline
\multicolumn{1}{|c|}{\multirow{3}{*}{\textbf{\begin{tabular}[c]{@{}c@{}}Compiler \\ Classification\end{tabular}}}}    & \multicolumn{2}{c|}{\textbf{Strands}}                                                                           & \multicolumn{3}{c|}{78.4}                                                                                     & \multicolumn{3}{c|}{\textbf{78.9}}                                                                            & \multicolumn{3}{c|}{59.3}                                                                                     & \multicolumn{3}{c|}{73.5}                                                                                     & \multicolumn{3}{c|}{70.6}                                                                \\ \cline{2-18} 
\multicolumn{1}{|c|}{}                                                                                                & \multicolumn{2}{c|}{\textbf{\begin{tabular}[c]{@{}c@{}}Basic\\ Blocks\end{tabular}}}                            & \multicolumn{3}{c|}{75.4}                                                                                     & \multicolumn{3}{c|}{\textbf{76.1}}                                                                                     & \multicolumn{3}{c|}{62.8}                                                                                     & \multicolumn{3}{c|}{71.3}                                                                                     & \multicolumn{3}{c|}{67.3}                                                                \\ \cline{2-18} 
\multicolumn{1}{|c|}{}                                                                                                & \multicolumn{2}{c|}{\textbf{Functions}}                                                                         & \multicolumn{3}{c|}{89.1}                                                                                     & \multicolumn{3}{c|}{\textbf{89.4}}                                                                            & \multicolumn{3}{c|}{72.9}                                                                                     & \multicolumn{3}{c|}{85.6}                                                                                     & \multicolumn{3}{c|}{81.3}                                                                \\ \hline
\multicolumn{1}{|c|}{\multirow{3}{*}{\textbf{\begin{tabular}[c]{@{}c@{}}Optimization\\ Classification\end{tabular}}}} & \multicolumn{2}{c|}{\textbf{Strands}}                                                                           & \multicolumn{3}{c|}{78.4}                                                                                     & \multicolumn{3}{c|}{\textbf{79.9}}                                                                            & \multicolumn{3}{c|}{70.5}                                                                                     & \multicolumn{3}{c|}{76.5}                                                                                     & \multicolumn{3}{c|}{73.9}                                                                \\ \cline{2-18} 
\multicolumn{1}{|c|}{}                                                                                                & \multicolumn{2}{c|}{\textbf{\begin{tabular}[c]{@{}c@{}}Basic\\ Blocks\end{tabular}}}                            & \multicolumn{3}{c|}{\textbf{69.6}}                                                                            & \multicolumn{3}{c|}{69.6}                                                                           & \multicolumn{3}{c|}{66.2}                                                                                     & \multicolumn{3}{c|}{67.3}                                                                                     & \multicolumn{3}{c|}{67.1}                                                                \\ \cline{2-18} 
\multicolumn{1}{|c|}{}                                                                                                & \multicolumn{2}{c|}{\textbf{Functions}}                                                                         & \multicolumn{3}{c|}{\textbf{75.4}}                                                                            & \multicolumn{3}{c|}{74.1}                                                                                     & \multicolumn{3}{c|}{67.4}                                                                                     & \multicolumn{3}{c|}{72.3}                                                                                     & \multicolumn{3}{c|}{68.9}                                                                \\ \hline
\multicolumn{3}{|c|}{\textbf{Strand Execution}}                                                                                                                                                                                         & \multicolumn{3}{c|}{\textbf{87.9}}                                                                            & \multicolumn{3}{c|}{86.5}                                                                                     & \multicolumn{3}{c|}{29.9}                                                                                     & \multicolumn{3}{c|}{37.0}                                                                                     & \multicolumn{3}{c|}{20.0}                                                                \\ \hline
\multicolumn{3}{|c|}{}                                                                                                                                                                                                                  & \multicolumn{1}{c}{\textit{Prec.}} & \multicolumn{1}{c}{\textit{Rec.}} & \multicolumn{1}{c|}{\textit{F1}}   & \multicolumn{1}{c}{\textit{Prec.}} & \multicolumn{1}{c}{\textit{Rec.}} & \multicolumn{1}{c|}{\textit{F1}}   & \multicolumn{1}{c}{\textit{Prec.}} & \multicolumn{1}{c}{\textit{Rec.}} & \multicolumn{1}{c|}{\textit{F1}}   & \multicolumn{1}{c}{\textit{Prec.}} & \multicolumn{1}{c}{\textit{Rec.}} & \multicolumn{1}{c|}{\textit{F1}}   & \multicolumn{1}{c}{\textit{Prec.}} & \multicolumn{1}{c}{\textit{Rec.}} & \textit{F1}   \\ \hline
\multicolumn{3}{|c|}{\textbf{Strand Recovery}}                                                                                                                                                                                          & \multicolumn{1}{c}{\textbf{96.9}}           & \multicolumn{1}{c}{\textbf{98.7}}          & \multicolumn{1}{c|}{\textbf{97.8}}          & \multicolumn{1}{c}{96.9}           & \multicolumn{1}{c}{98.4}          & \multicolumn{1}{c|}{97.6}          & \multicolumn{1}{c}{65.7}           & \multicolumn{1}{c}{29.9}          & \multicolumn{1}{c|}{41.1}          & \multicolumn{1}{c}{81.1}           & \multicolumn{1}{c}{75.5}          & \multicolumn{1}{c|}{79.2}          & \multicolumn{1}{c}{80.1}           & \multicolumn{1}{c}{56.7}          & 66.3          \\ \hline
\end{tabular}%
}
\caption{Extrinsic evaluation results.}
\label{table:extrinsic_res}
\end{table*}

% !TEX root =  main.tex
\section{Related Works}

Binary code representation techniques can be subdivided into two main branches: manual features selection \cite{malClass13Liang, drebin2014arp, bov2015padmana,gemini, punstrip20evans, XFL21evans}  and unsupervised features extraction. Since our paper proposes an assembly model we discuss works that use or propose instruction embedding models. We first focus on the instruction embedding models proposed in the literature, categorising them according to the defining properties identified in the Background Section \ref{sec:back}:  the distributed representation learning used, the preprocessing of assembly instructions, and the extraction methodology for assembly sequences. Finally, we discuss how these models have been used to solve binary analysis tasks. 

\subsection{Distributed Representation Learning} The distributed representation learning technique is the neural architecture and the training tasks used by the instruction models to create useful embedding vectors.  
The most common is word2vec \cite{w2v}. Word2vec has been used, with minimal modifications, by Eklavya \cite{eklavya}, SAFE \cite{safe19mass} and others \cite{nmt19Zuo, deepbindiff, InvestigatingGE19mass}. A notable difference is  Asm2vec \cite{asm2vec} which uses PV-DM \cite{pvdm}, a variation of word2vec that simultaneously creates instructions and function embeddings. We remark that PV-DM cannot be fine-tuned. PalmTree \cite{palmtree2021li}, that we described in Section \ref{subsec:palmtree}, uses a transformer. 

\subsection{Preprocessing of Assembly Instructions}
The preprocessing is characterised by the substitution policy for information in the raw assembly instruction and the tokenization policy. We classify the substitution policies in {\em aggressive} or {\em light}. In an aggressive policy, lots of information contained in the assembly instructions are removed or changed. 
An example is DeepBinDiff \cite{deepbindiff} that replaces all constants and pointers with special tokens and renames registers according to their lengths in bytes (e.g. \asm{ecx} becomes \asm{reg4}). InnerEye \cite{nmt19Zuo}, instead, replaces all constants, strings, and function names with special symbols. BinDeep \cite{bindeep} substitutes operands based on predefined categories (e.g. general register, direct memory reference, etc.). All the above policies are aggressive; indeed, InnerEye \cite{nmt19Zuo} wastes fundamental information that a library function call could bring and DeepBinDiff \cite{deepbindiff} loses register names that could be relevant, for instance, to understand the data flow. 

Light preprocessing policies are applied by SAFE \cite{safe19mass} and PalmTree \cite{palmtree2021li} which keep small constant values and replace values above a predefined threshold with a special token. SAFE \cite{safe19mass} replaces all memory addresses with the same token, PalmTree uses different tokens for generic memory locations and the ones pointing to strings.

Regarding tokenization, some works consider an entire instruction as a token \cite{safe19mass} or split an instruction into opcode and operands \cite{asm2vec}, while others use a word-based tokenizer by splitting instructions on symbols (e.g. spaces or special characters) \cite{trex}, \cite{palmtree2021li}. The latter is a fine-grained tokenization strategy that is useful to reduce the vocabulary size and allow the upstream network to separately learn the semantics of each token (mnemonics, registers, etc.). No one used automatic tokenization strategies like WordPiece \cite{wordpiece}.

\subsection{Binary Analysis Solutions using Embedding Models}
Deep-learning-based solutions can be categorized into approaches using Graph Neural Networks (GNN), Recurrent Neural Networks (RNN), and Transformer architecture. \cite{InvestigatingGE19mass, nero20david} use a GNN applied to function control flow graphs after transforming each block into a vector representation; this transformation is done by aggregating the instruction embeddings of each block. The resulting architecture is applied to the binary similarity\cite{InvestigatingGE19mass}, the compiler provenance problem \cite{InvestigatingGE19mass}, function naming problem \cite{nero20david}. 
Eklavya \cite{eklavya} and InnerEye \cite{nmt19Zuo} are examples of RNN-based solutions applied to the recovery of arguments used by a binary function \cite{eklavya}, and basic-block similarity\cite{nmt19Zuo}. Another work is SAFE \cite{safe19mass} which added a self-attention layer on top of an LSTM to solve the binary similarity problem. 
Among transformer-based solutions, we can find \cite{innomine}, \cite{nero20david} that apply transformers encoder-decoder architecture to recover function names from stripped binaries. Other works \cite{orders20Yu, trex} use transformer encoder for the binary similarity problem. In particular, \cite{orders20Yu} uses a transformer encoder to obtain basic blocks embeddings to be fed into a GNN which is trained in conjunction with a Convolutional Neural Network (CNN) to produce a function embedding for binary similarity. Unfortunately, \cite{orders20Yu}  does not release the code and the paper lacks of the details needed to reimplement their proposal. Trex \cite{trex} pre-trains a transformer encoder on function micro-traces and then uses this pre-trained model to produce embeddings for function similarity.

\section{Conclusion}
We presented BinBert, an execution-aware assembly language model. BinBert is trained on a big dataset of assembly strands and symbolic expressions. BinBert has shown state-of-the-art performance highlighting the relevance of execution-awareness. Our evaluation shows that BinBert is an encoder model that can be used to solve several tasks related to binary analysis using a fine-tune dataset of relatively small size. 
The generality of the model is an important strength and we believe that BinBert can be fruitfully applied to other downstream tasks as encoder layer of complex neural networks.

\bibliographystyle{IEEEtranS}
\bibliography{IEEEabrv, bibliography}
% !TEX root =  main.tex
\section{Appendix}

\subsection{Opcode an Operand Classes}
\label{appendix:op_class}
\begin{figure*}
    \center
    \includegraphics[width=\textwidth]{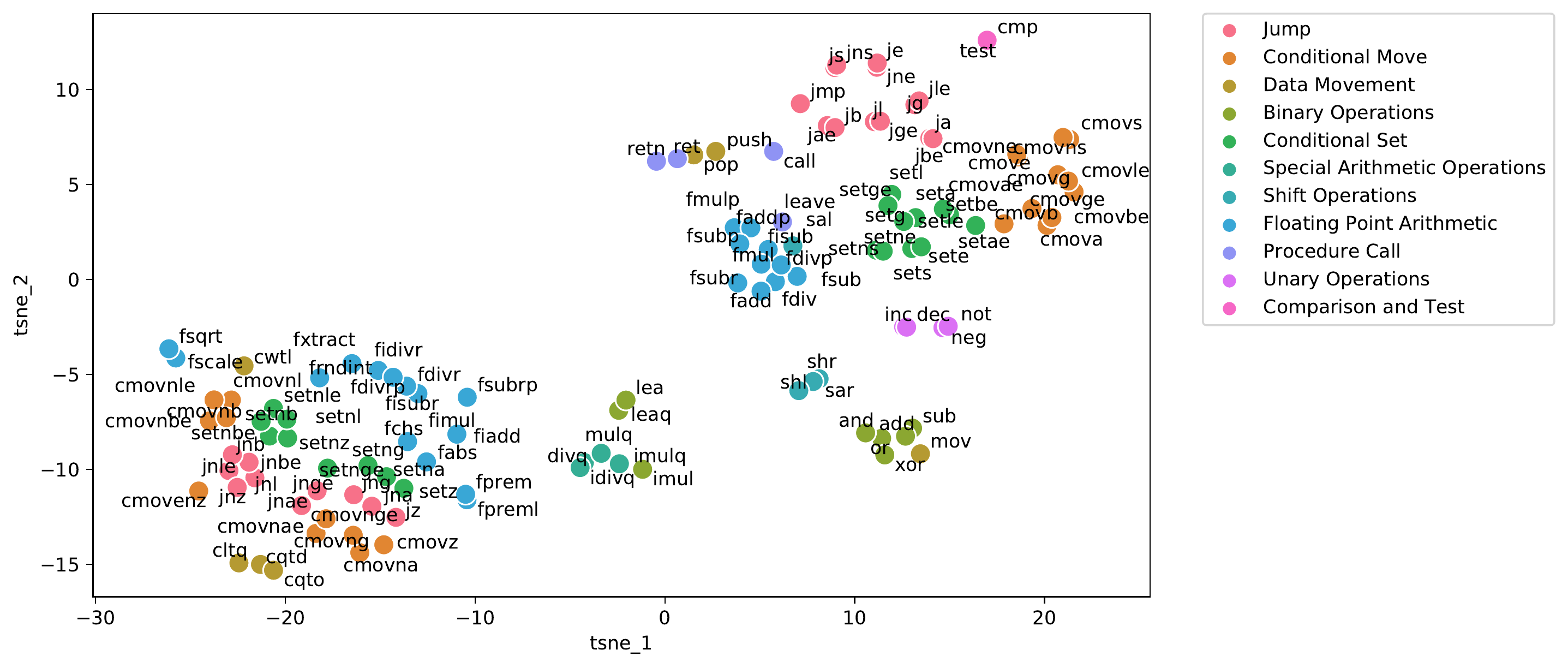}
    \caption{Visualization of opcode embeddings in a two-dimensional space with t-SNE.}
\label{fig:tsne}
\end{figure*}

In Table \ref{table:opcode_types} and \ref{table:operand_types} we reported the classes that we have used to categorise instructions for the opcode and operand outlier detection tasks respectively. 
We performed a qualitative analysis based on the visualization of the clusters of opcodes learned by BinBert. To do so, we first used BinBert to convert each opcode into a vector, and we then applied t-SNE \cite{tsne} to visualize opcodes in a two-dimensional space. Results are shown in Figure \ref{fig:tsne}. Opcodes are well clustered according to their semantics. Specifically, we can identify two symmetric regions in which jumps, conditional move, and conditional set are split based on the condition checked (either positive or negative). We can also identify a region containing operations performing multiplications (\asm{imul}, \asm{lea}) and divisions (\asm{divq}), and other arithmetic operations (\asm{add}, \asm{sub}). Another interesting example is given by the region containing \asm{ret}, \asm{pop}, \asm{push} and \asm{call}: they all manipulate the stack. 

\begin{table}[]

\center
\begin{tabular}{|c|c|}
\hline
\textbf{Type}                                                             & \textbf{Opcodes}                                                                                                                                                                                                                                                                           \\ \hline
Data Movement                                                             & \begin{tabular}[c]{@{}c@{}}\asm{mov}, \asm{push}, \asm{pop}, \asm{cwtl}, \\ \asm{cltq}, \asm{cqto},\asm{cqtd}\end{tabular}                                                                                                                                                                                             \\ \hline
Unary Operations                                                          & \asm{inc}, \asm{dec}, \asm{neg}, \asm{not}                                                                                                                                                                                                                                                                 \\ \hline
Binary Operations                                                         & \begin{tabular}[c]{@{}c@{}}\asm{add}, \asm{sub},\asm{imul}, \asm{xor}, \\ \asm{or}, \asm{and}, \asm{lea}, \asm{leaq}\end{tabular}                                                                                                                                                                                          \\ \hline
Shift Operations                                                          & \asm{sal}, \asm{sar}, \asm{shr}, \asm{shl}                                                                                                                                                                                                                                                                 \\ \hline
\begin{tabular}[c]{@{}c@{}}Special Arithmetic \\ Operations\end{tabular}  & \begin{tabular}[c]{@{}c@{}}\asm{imulq}, \asm{mulq}, \asm{idivq}, \\ \asm{divq}\end{tabular}                                                                                                                                                                                                                \\ \hline
\begin{tabular}[c]{@{}c@{}}Comparison and Test\\ Intructions\end{tabular} & \asm{cmp}, \asm{test}                                                                                                                                                                                                                                                                              \\ \hline
\begin{tabular}[c]{@{}c@{}}Conditional Set\\ Instructions\end{tabular}    & \begin{tabular}[c]{@{}c@{}}\asm{sete}, \asm{setz}, \asm{setne}, \\ \asm{setnz}, \asm{sets},\asm{setns}, \asm{setg}, \\ \asm{setnle},\asm{setge}, \asm{setnl}, \asm{setl}, \\ \asm{setnge},\asm{setle}, \asm{setng}, \asm{seta}, \\ \asm{setnbe}, \asm{setae}, \asm{setnb}, \asm{setbe}, \\ \asm{setna}\end{tabular}                                                        \\ \hline
Jump Instructions                                                         & \begin{tabular}[c]{@{}c@{}}\asm{jmp}, \asm{je}, \asm{jz}, \asm{jne}, \asm{jnz}, \\ \asm{js}, \asm{jns}, \asm{jg}, \asm{jnle},\asm{jge}, \\ \asm{jnl}, \asm{jl}, \asm{jnge}, \asm{jle}, \asm{jng}, \\ \asm{ja}, \asm{jnbe}, \asm{jae},\asm{jnb}, \asm{jb}, \\ \asm{jnae}, \asm{jbe}, \asm{jna}\end{tabular}                                                                             \\ \hline
\begin{tabular}[c]{@{}c@{}}Conditional Move\\ Instructions\end{tabular}   & \begin{tabular}[c]{@{}c@{}}\asm{cmove}, \asm{cmovz}, \asm{cmovne}, \\ \asm{cmovenz}, \asm{cmovs}, \asm{cmovns},\\ \asm{cmovg}, \asm{cmovnle}, \asm{cmovge}, \\ \asm{cmovnl}, \asm{cmovnge}, \asm{cmovle},\\ \asm{cmovng}, \asm{cmova}, \asm{cmovnbe},\\  \asm{cmovae}, \asm{cmovnb}, \asm{cmovb}, \\ \asm{cmovnae}, \asm{cmovbe}, \asm{cmovna}\end{tabular}                    \\ \hline
\begin{tabular}[c]{@{}c@{}}Procedure Call\\ Instructions\end{tabular}     & \asm{call}, \asm{leave}, \asm{ret}, \asm{retn}                                                                                                                                                                                                                                                             \\ \hline
\begin{tabular}[c]{@{}c@{}}Floating Point\\ Arithmetic\end{tabular}       & \begin{tabular}[c]{@{}c@{}}\asm{fabs}, \asm{fadd}, \asm{faddp}, \asm{fchs}, \\ \asm{fdiv}, \asm{fdivp}, \asm{fdivr},\asm{fdivrp}, \\ \asm{fiadd}, \asm{fidivr},\asm{fimul}, \asm{fisub}, \\ \asm{fisubr}, \asm{fmul},\asm{fmulp}, \asm{fprem}, \\ \asm{fpreml}, \asm{frndint}, \asm{fscale}, \asm{fsqrt}, \\ \asm{fsub}, \asm{fsubp}, \asm{fsubr}, \\ \asm{fsubrp}, \asm{fxtract}\end{tabular} \\ \hline
\end{tabular}%

\caption{Opcode classes used to categorise instructions for the opcode outlier detection task.}
\label{table:opcode_types}
\end{table}

\begin{table}
%\resizebox{\linewidth}{!}{%
\center
\begin{tabular}{|c|c|c|c|}
\hline
\textbf{Type} & \textbf{Operand 1} & \textbf{Operand 2} & \textbf{Operand 3} \\ \hline
none          & -                  & -                  & -                  \\ \hline
cnst          & immediate         & -                  & -                  \\ \hline
reg           & register           & -                  & -                  \\ \hline
ref           & memory             & -                  & -                  \\ \hline
reg-reg       & register           & register           & -                  \\ \hline
reg-cnst      & register           & immediate         & -                  \\ \hline
reg-ref       & register           & memory             & -                  \\ \hline
ref-reg       & memory             & register           & -                  \\ \hline
ref-cnst      & memory             & immediate         & -                  \\ \hline
tri           & any                & any                & any                \\ \hline
\end{tabular}%
\caption{Operand classes used to categorise instructions for the operand outlier detection task.}
\label{table:operand_types}
\end{table}

\subsection{Figures for Compiler Provenance at Function Level}\label{appendix:confusionmatrices}
\begin{figure*}
\centering
\subfloat[Confusion matrix for the compiler classification task at function level with BinBert. \label{fig:func_compiler} ]{\includegraphics[width=.45\textwidth]{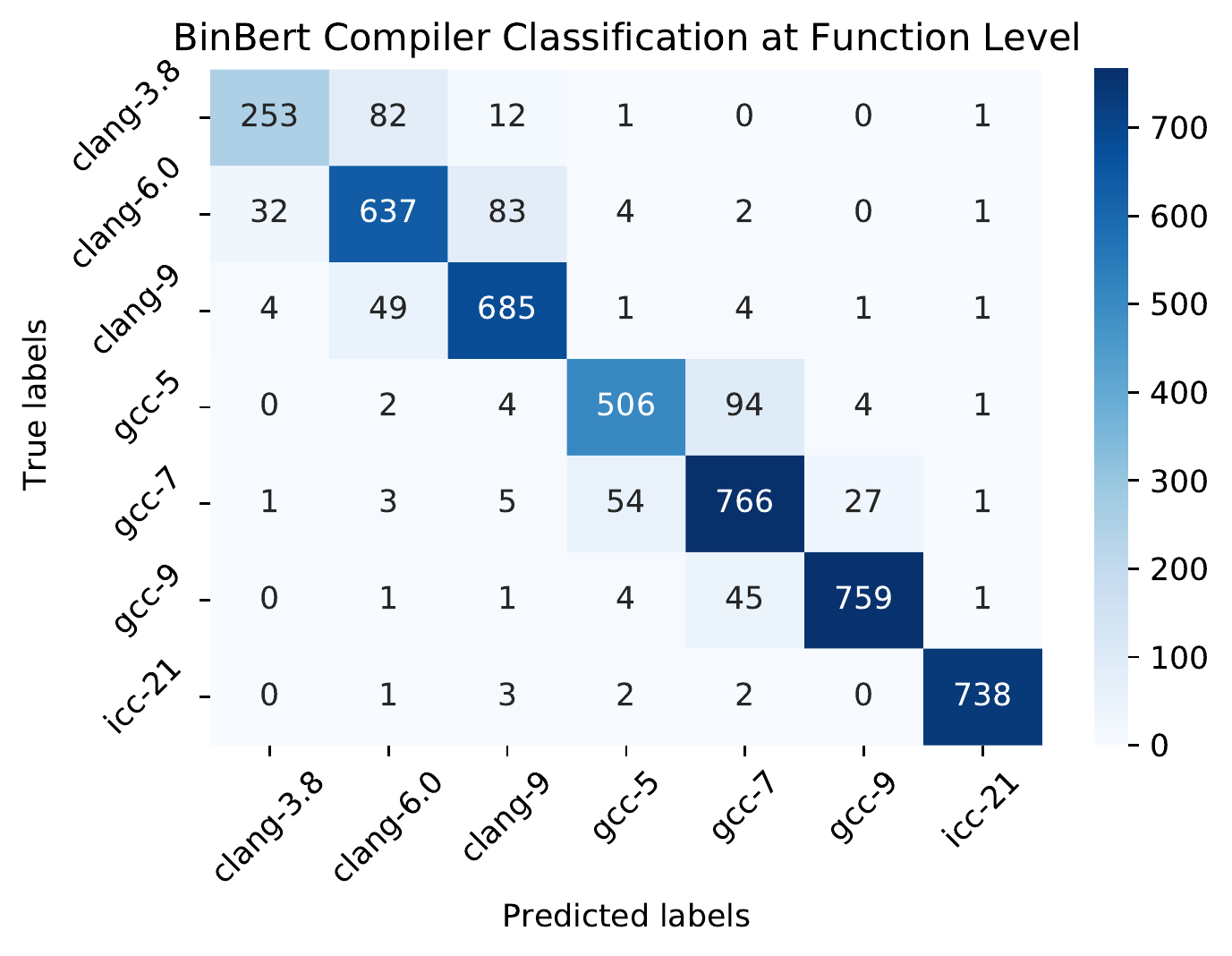}}\hfill
\subfloat[Confusion matrix for the optimization classification task at function level with BinBert. \label{fig:func_opt} ]{\includegraphics[width=.45\textwidth]{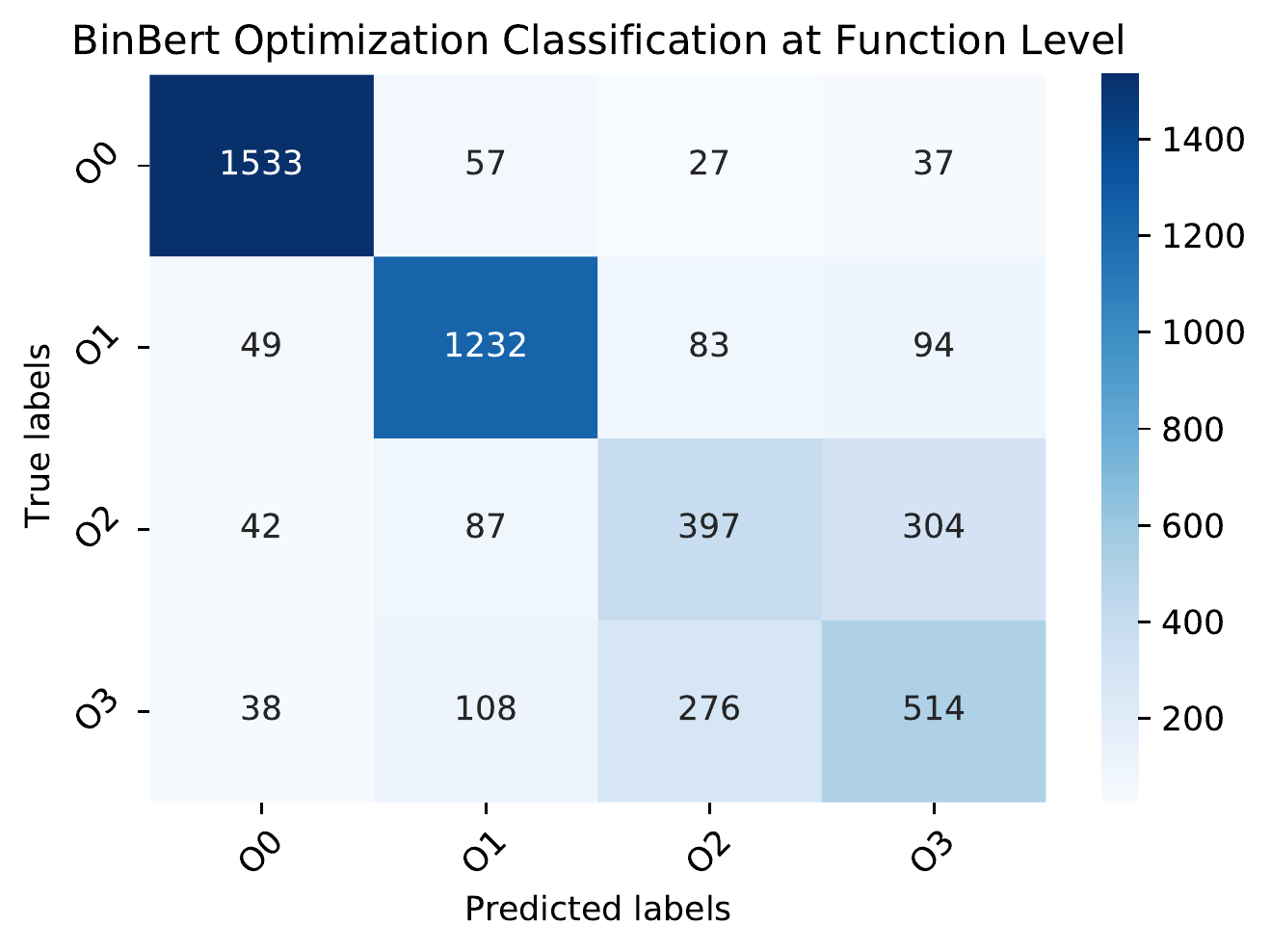}}\hfill
\caption{  Results for the {\bf extrinsic compiler provenance} task. \label{fig:comp_prov}}
\end{figure*}

In Figure \ref{fig:comp_prov} we reported the confusion matrices obtained by using BinBert on the compiler and optimization classification tasks. We can observe that BinBert can clearly distinguish among different compiler families and it only gets confused with different versions in the same family. Similar behavior can be observed in the optimization classification task. It is easier for BinBert to distinguish optimized (O1, O2, O3) vs unoptimized code (O0) than to recognize the specific optimization level used to compile it.

\subsection{Figures for Extrinsic and Intrinsic Block Similarity}\label{appendix:similarity}
\begin{figure}
\centering
\subfloat[Precision for the  top-$k$ answers with $k \leq 30$. \label{fig:precision} ]{\includegraphics[width=.45\textwidth]{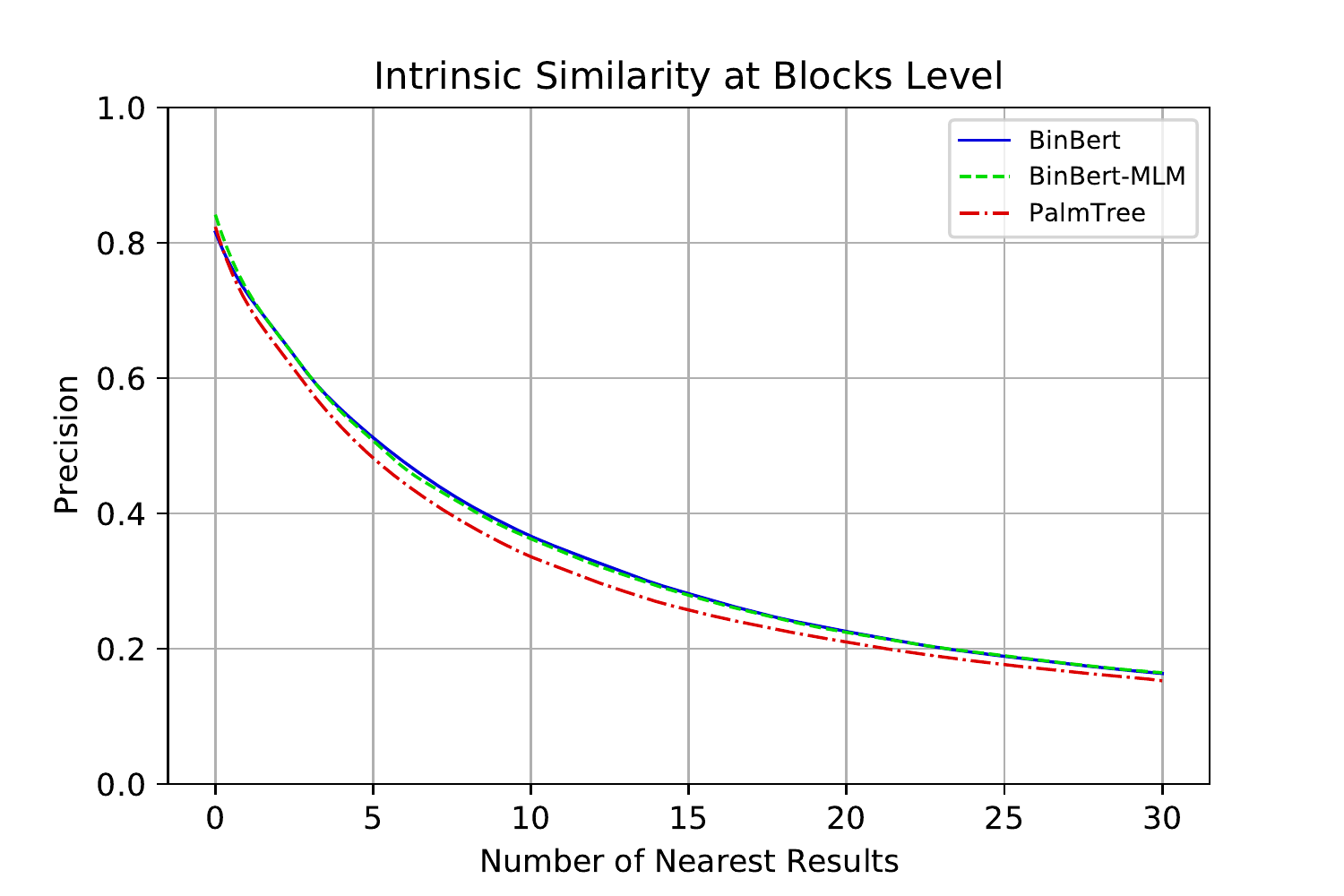}}\hfill
\subfloat[nDCG for the top-$k$ answers with $k \leq 30$. \label{fig:ndcg} ]{\includegraphics[width=.45\textwidth]{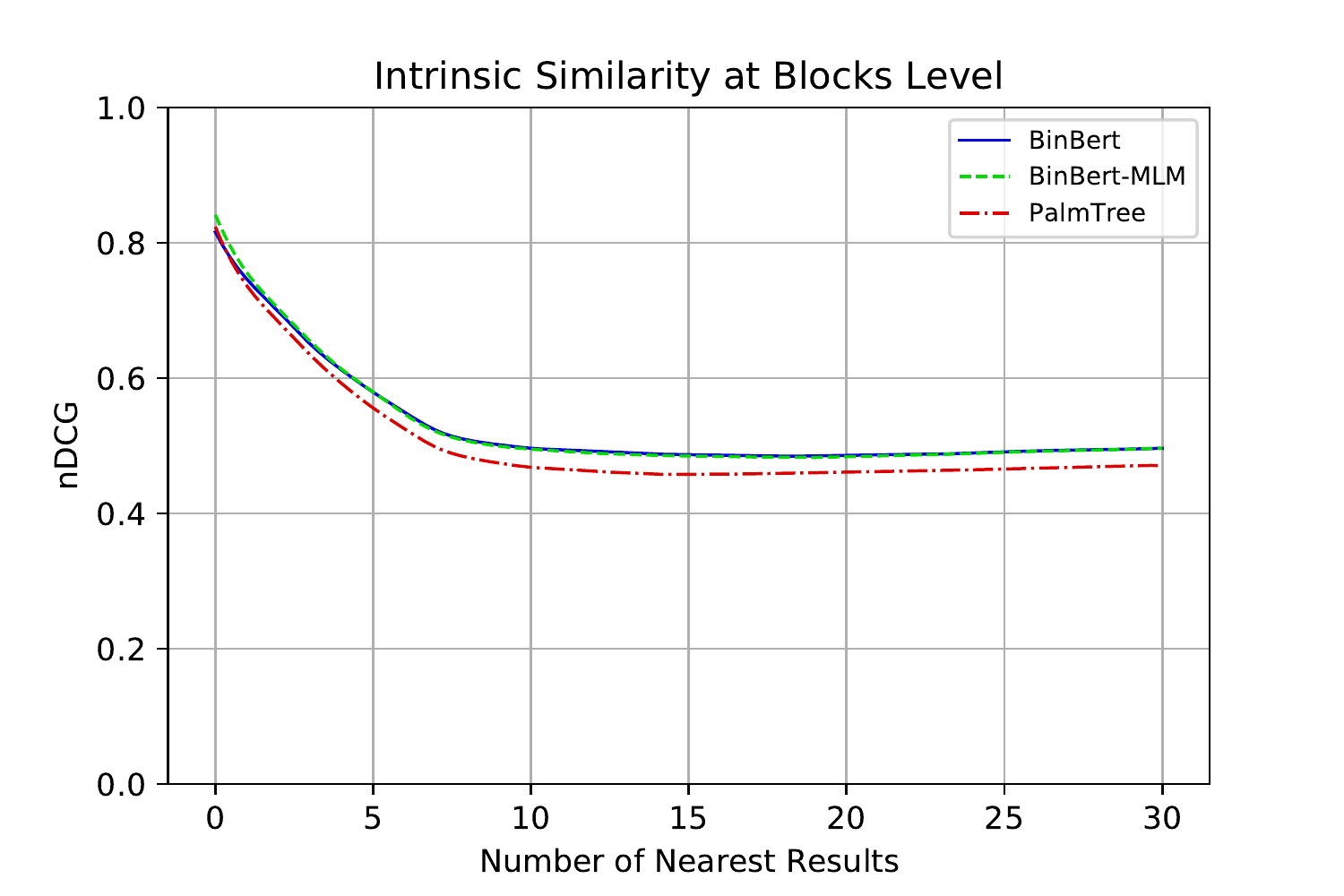}}\hfill
\subfloat[Recall for the  top-$k$ answers with $k \leq 30$. \label{fig:recall} ]{\includegraphics[width=.45\textwidth]{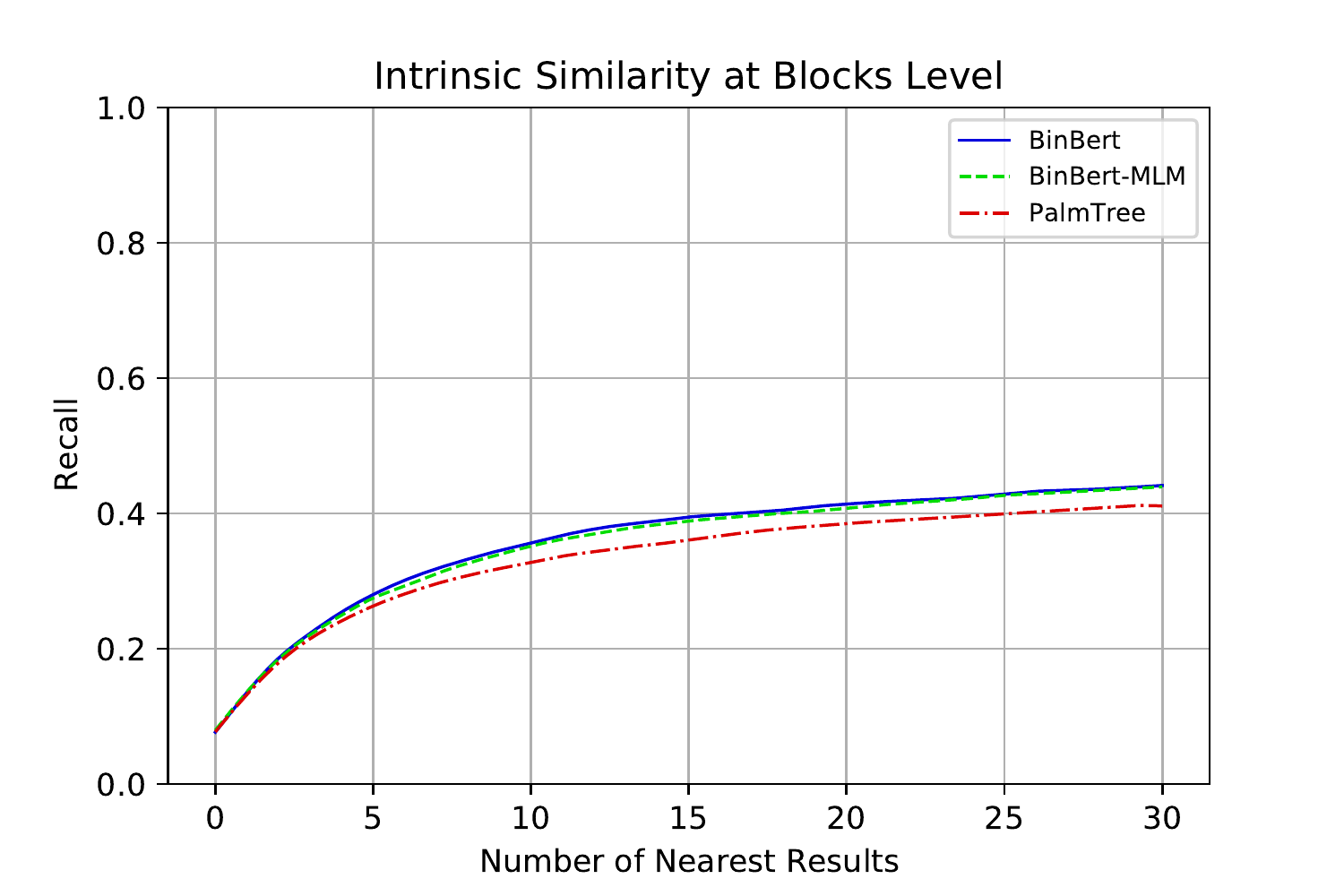}}\hfill
\caption{  Results for the {\bf intrinsic block similarity} task. Database of 6460 CFG basic blocks, average on 505 queries. \label{fig:blocksearch_intrinsic}}
\end{figure}

\begin{figure}
\centering
\subfloat[Precision for the  top-$k$ answers with $k \leq 30$. \label{fig:precision} ]{\includegraphics[width=.45\textwidth]{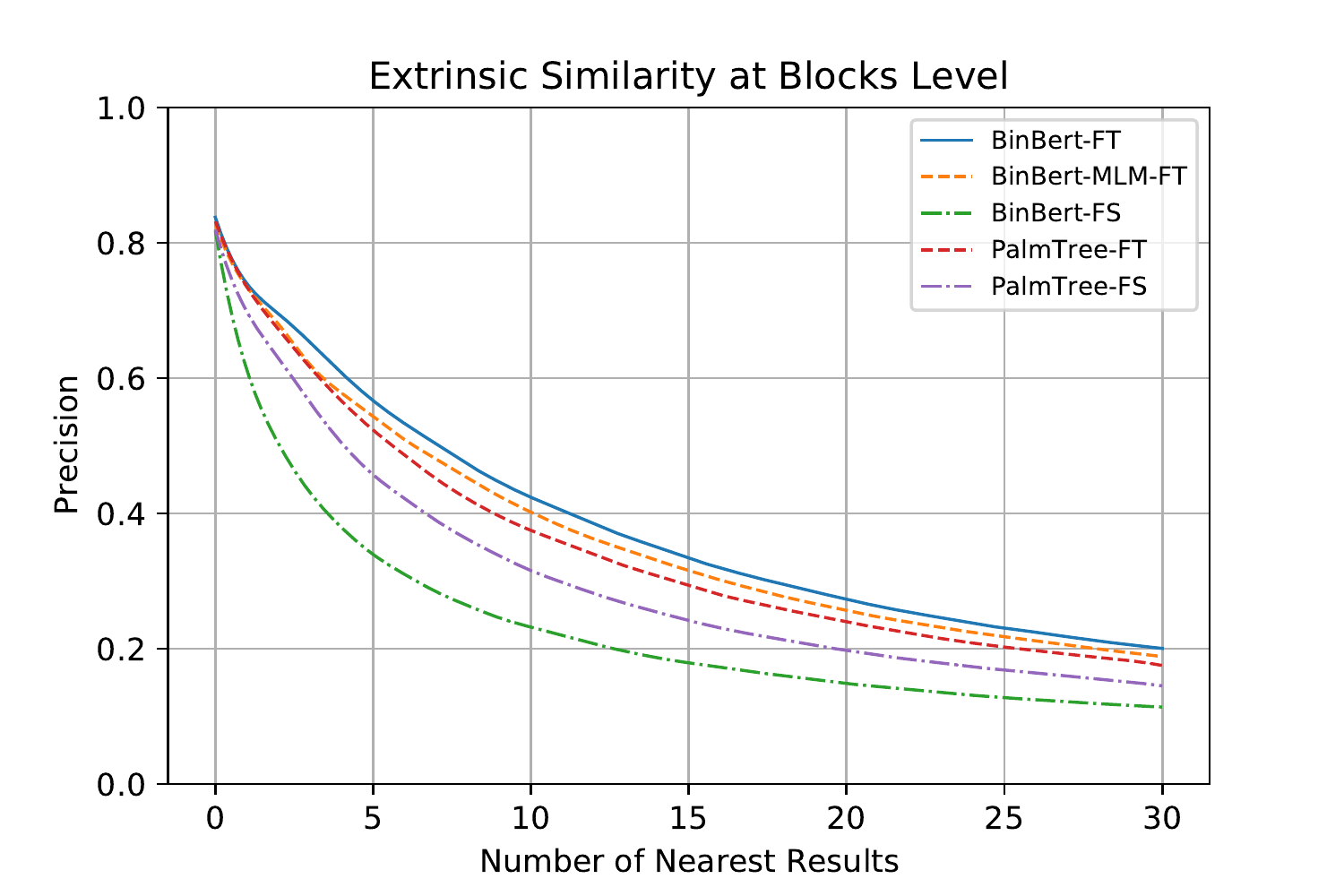}}\hfill
\subfloat[nDCG for the top-$k$ answers with $k \leq 30$. \label{fig:ndcg} ]{\includegraphics[width=.45\textwidth]{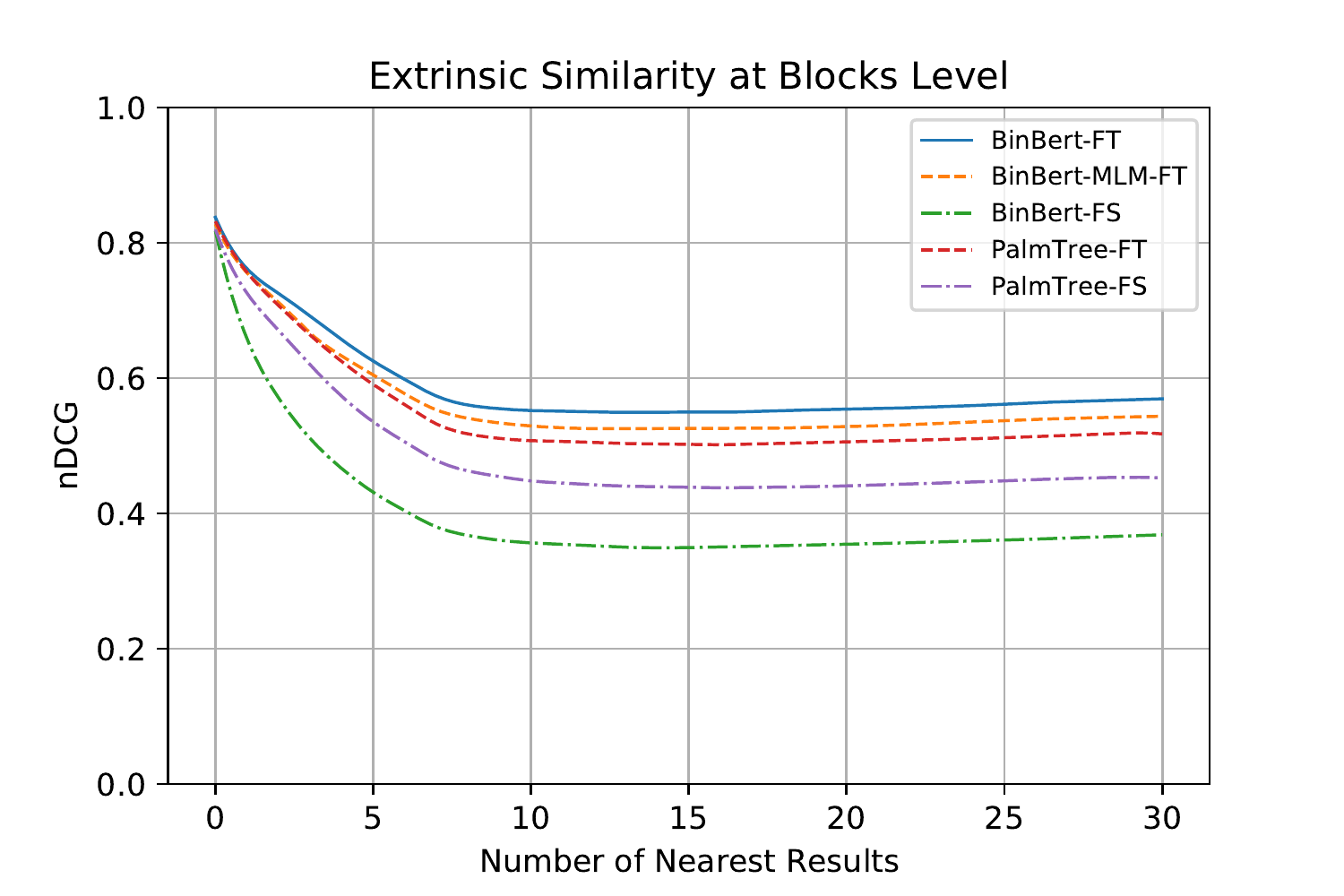}}\hfill
\subfloat[Recall for the  top-$k$ answers with $k \leq 30$. \label{fig:recall} ]{\includegraphics[width=.45\textwidth]{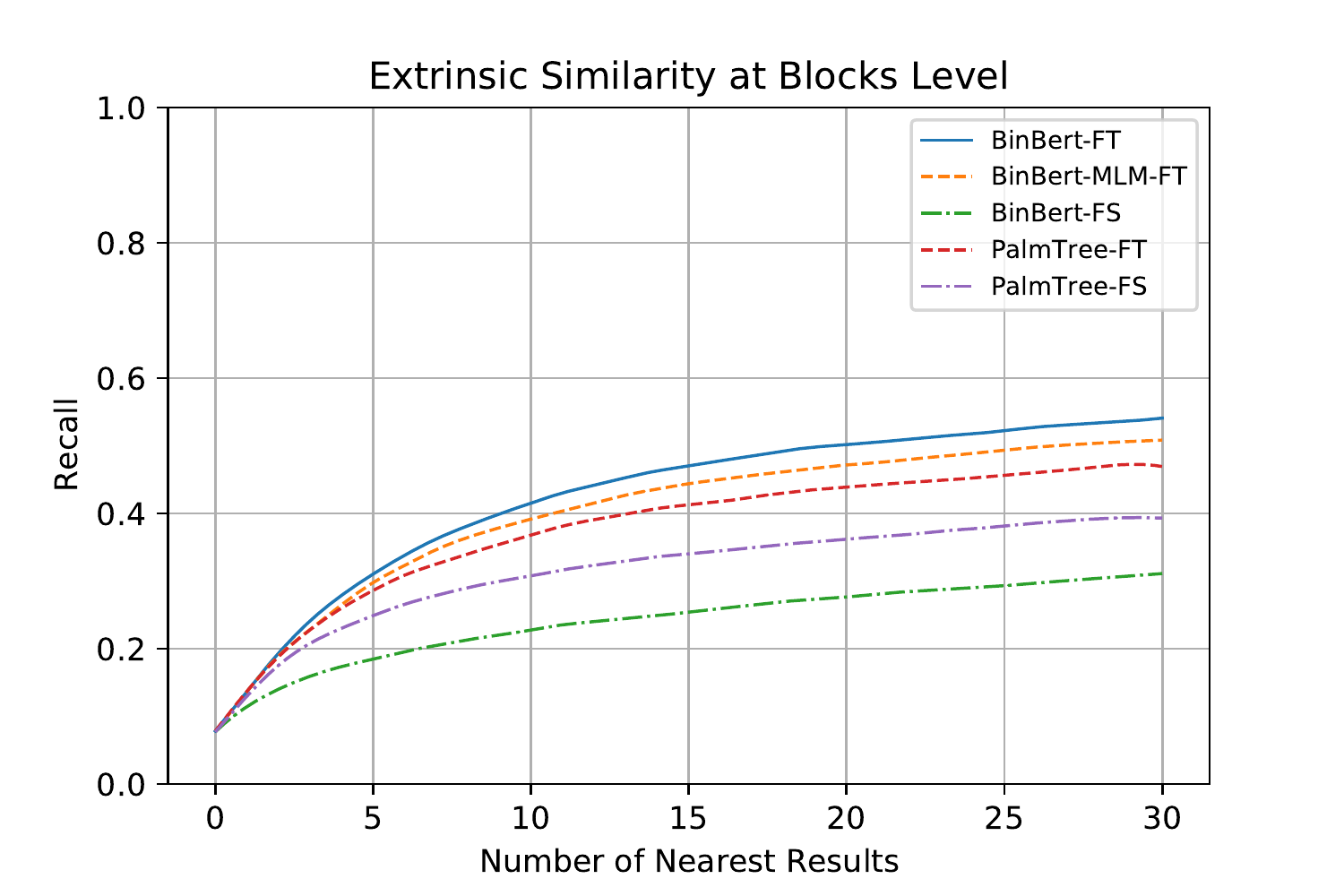}}\hfill
\caption{  Results for the {\bf extrinsic block similarity} task. Database of 6460 CFG basic blocks, average on 505 queries. \label{fig:blocksearch_extrinsic}}
\end{figure}

We reported the results of the intrinsic and extrinsic block similarity tasks in Figures \ref{fig:blocksearch_intrinsic} and \ref{fig:blocksearch_extrinsic}. BinBert shows slightly higher performances with respect to PalmTree at the block level.

\end{document}